\documentstyle{mn}

\input{epsf.sty}

\def\mpc {h^{-1} {\rm{Mpc}}}
\def\and  {\it {et al.} \rm}
\def\rmd {\rm d}

\def\etal{{\rm et~al. }}
\def\spose#1{\hbox to 0pt{#1\hss}}
\def\simlt{\mathrel{\spose{\lower 3pt\hbox{$\mathchar"218$}}
     \raise 2.0pt\hbox{$\mathchar"13C$}}}
\def\simgt{\mathrel{\spose{\lower 3pt\hbox{$\mathchar"218$}}
     \raise 2.0pt\hbox{$\mathchar"13E$}}}
\def\beq{\begin{equation}}
\def\eeq{\end{equation}}
\def\bce{\begin{center}}
\def\ece{\end{center}}
\def\bea{\begin{eqnarray}}
\def\eea{\end{eqnarray}}
\def\ben{\begin{enumerate}}
\def\een{\end{enumerate}}

\def\nn{\nonumber}

\def\brr{\begin{array}}
\def\err{\end{array}}

\def\etal{{\rm et~al. }}
\def\hmpc{\;h^{-1}{\rm Mpc}}

\def\kms{{\rm \;km\;s^{-1}}}

\def\kmsmpc{\kms\;{\rm Mpc}^{-1}}
\def\lya{Ly$\alpha$}

\def\nh1{n_{\rm HI}}

\def\K{{\rm K}}

\def\p1dk{P_{\rm 1D}(k)}
\def\simlt{\mathrel{\spose{\lower 3pt\hbox{$\mathchar"218$}}
     \raise 2.0pt\hbox{$\mathchar"13C$}}}
\def\simgt{\mathrel{\spose{\lower 3pt\hbox{$\mathchar"218$}}
     \raise 2.0pt\hbox{$\mathchar"13E$}}}

\newcommand{\xibar}{\overline{\xi}}

\def\Or{{\cal O}}

\def\calG{{\cal G}}

\def\lya{Ly$\alpha$}

\font\twelveBF=cmmib10 scaled 1200

\newcommand{\x}{\hbox{\twelveBF x}}

\newcommand{\de}{\delta}

\begin{document}

\title[Predictions for the clustering of the 
Lyman-alpha Forest]
{Predictions for the clustering properties of the  
Lyman-alpha Forest - I. One-point statistics}

\author[E. Gazta\~{n}aga and R.A.C. Croft]{
 Enrique Gazta\~{n}aga$^{1}$
and Rupert A. C. Croft$^{2}$
\vspace{1mm}\\
$^1$ 
Consejo Superior de Investigaciones Cient\'{\i}ficas (CSIC), 
Institut d'Estudis Espacials de Catalunya (IEEC), \\
Edf. Nexus-201 - c/ Gran Capitan 2-4, 08034 Barcelona, SPAIN \\
$^2$ Astronomy Department, Harvard University, 60 Garden St,
Cambridge, MA 01238, USA}

\maketitle 
 
\def\mpc {h^{-1} {\rm Mpc}}
\def\impc {h {\rm Mpc}^{-1}}
\def\and  {{\it {et al.} }}
\def\rmd {{\rm d}}

\begin{abstract}

We present predictions for the one-point probability distribution and 
cumulants of the transmitted QSO flux in the high redshift 
Lyman-$\alpha$ Forest. 
We make use of the  
correlation between the 
Lyman-$\alpha$ optical depth and the underlying matter density
predicted by gravitational instability theory 
and seen in numerical hydrodynamic simulations. We have modelled the growth 
of matter fluctuations using the non-linear shear-free dynamics,
an approximation which reproduces 
well the results of perturbation theory for the cumulants  in the 
linear and weakly non-linear clustering regime.
As high matter overdensities tend to saturate in spectra, 
the statistics of the flux distribution are dominated by weakly 
non-linear overdensities. As a result, our analytic 
approach can produce accurate predictions,
when tested against  N-body simulation results, 
even when the underlying matter field has
rms fluctuations larger than unity. 
Our treatment can be applied to either
Gaussian or non-Gaussian initial conditions. Here we 
concentrate  on the former case, but also include a study of a
specific non-Gaussian model.
We discuss how the methods and predictions we present
can be used as a tool to study the
generic clustering properties of the \lya\ forest at high-redshift.
 With such an approach, rather than concentrating
on simulating specific cosmological models, we may be in the position to
directly test our assumptions for the Gaussian nature of the initial 
conditions, and the gravitational instability origin of structure itself.    
In a separate paper we present results for two-point
statistics.

\end{abstract}


\section{Introduction}

The \lya\ forest
(Lynds 1971, Sargent \etal 1980, see Rauch 1998 for a review)
 arises naturally in cosmological structure formation scenarios
where gravitational instability acts on small initial density perturbations.
In hydrodynamic simulations 
  of  such models
(Cen \etal 1994, Zhang \etal 1995, Hernquist \etal 1996, 
Wadsley \& Bond 1996, Theuns \etal 1998, see also the analytical modelling
 of e.g., Bi 1993, Reisenegger and Miralda-Escud\'{e} 1995),
 most of the absorption seen in high redshift
QSO spectra is generated by residual neutral hydrogen in
a continuous fluctuating photoionized
intergalactic medium.
In such a picture, absorbing structures have a large physical
extent. Observational support for this has come from 
comparison of the \lya\ forest in adjacent QSO lines of sight (Bechtold \etal
1994, Dinshaw \etal 1994, 1995, Crotts \& Fang 1998).
 For matter in this phase, it is predicted and  found in simulations that  the 
 underlying mass density field at a particular point
can be related to the optical depth for \lya\ absorption 
(see e.g., Croft \etal 1997)
and hence a directly observable quantity, the transmitted flux in the QSO
 spectrum.

 Much work has been devoted to studying
the statistical properties of the mass density field and the generic
predictions of the gravitational instability picture.
With the 
\lya\ forest as a probe of the density,
 we avoid many of the uncertainties associated with the use of the 
galaxy distribution to test theories.
In principle, it should be possible, by combining our theoretical knowledge
of gravitational clustering with observations of \lya\ absorption, to 
test the Gaussianity of the initial density field, the picture
of \lya\ formation, and the gravitational instability scenario itself.
In this paper, we will concentrate on one-point statistics, namely the 
one point probability distribution function (PDF)
 of the transmitted flux, and its moments.
We will use as a tool the spherical collapse or shear-free
model for the 
evolution of density perturbations which Fosalba \& Gazta\~{n}aga
(1998a hereafter FG98, 1998b)  have shown to be a good 
approximation to the growth of clustering in the 
weakly non-linear regime
and which we find 
works well in the  density regime appropriate to the study of the
 \lya\ forest. Two point statistics, which probe the scale dependence of 
clustering will be examined in an accompanying paper (
Gazta\~{n}aga \& Croft 1999, Paper II).

From  observations  of the \lya\ forest, 
 we can measure the PDF
 of the flux and its moments.
 The high resolution spectra of the forest taken by the Keck
telescope (Hu \etal 1995, Lu \etal 1996,
Rauch \etal 1997,
Kirkman \& Tytler 1997,
Kim \etal 1997) 
allow us to resolve structure in the flux 
distribution, and  make high precision, shot noise-free
 measurements of these flux statistics.
 Here we will use 
the statistical properties of the matter distribution,
$\rho(x)$, to predict these 
observable quantities.

There are a number of studies which predict the evolution of the clustering
of density fluctuations, and in particular of the  PDF.
The Zel'dovich Approximation (ZA) was used by Kofman et al. (1994). 
Althought the ZA reproduces important aspects of  non-linear
dynamics, it only results in
 a poor approximation to the PDF and its moments.
This can be quantified by noticing, for example, that the hierarchical
skewness $S_3=\xibar_3/\xibar_2^2$ 
in the ZA is $S_3=4$ (at leading order in $\xibar$) instead
 of the Perturbation Theory (PT) result
$S_3=34/7$ (see e.g., Peebles 1980).
 One way to improve on this is to use the  PT cumulants
to derive the PDF from the Edgeworth
expansion (Juszkiewicz \etal 1995, Bernardeau \& Kofman 1995).
In this case the PDF is predicted to an 
accuracy given by order of the cumulants involved.
 Protogeros \& Scherrer (1997)
introduced the use of a 
 local Lagrangian mapping (that relates the initial
and evolved fluctuation) as a generic way to 
predict the PDF. In this case, the PDF is obtained simply by
applying a change of
variables (the mapping) to the PDF of the initial conditions.
The best of these two approaches is obtained when the Lagrangian mapping
is taken to be that of spherical collapse (FG98), which 
recovers the PT cumulants to arbitrary order in the weakly 
non-linear regime. There is yet another possibility, 
which involves performing a perturbative expansion and
directly relating the moments of the flux to the moments of the mass
(along the lines proposed in a different context by Fry \& Gaztanaga 1993). 
This  approach does not use the density PDF, 
and could incorporate more exact calculations
for the (non-linear) density  moments.

Our plan for this paper is as follows.
In Section 2 we  outline the physical basis for the relation we adopt 
between \lya\ optical depth and the mass distribution.
In Section 3 we describe our model for following the evolution
of the PDF of the density and flux, using non-linear mapping
relations presented in appendix A1. The cumulants of the flux distribution
predicted by fully non-linear dynamics
are described in Section 4, together with
the predictions of perturbation theory. The modelling
of the effects of redshift distortions and thermal broadening
is also described. In Section 5, we compare our analytical results to those
measured from simulated spectra, generated using N-body simulations.
In Section 6, we discuss the effects of non-Gaussian initial conditions,
the redshift evolution of the one-point flux statistics, and the bias between
flux and mass fluctuations. We also compare to other work on
the statistics of the \lya\ forest flux. Our summary and conclusions
form Section 7. 

\section{Lyman-alpha absorption and its relation to the mass
distribution}

As mentioned in Section 1, the model we use to relate
 \lya\ absorption to the distribution of mass is motivated by 
the results of numerical simulations which solve the full equations of 
hydrodynamics and gravity, some including star formation in high density
 regions. It was found in these simulations 
(e.g., Hernquist \etal 1996) that most of the volume of the 
Universe at high redshift ($z \simgt2$, see Dav\'{e} \etal 1999
for the situation at later times)
 is filled with a warm ($10^{4}$ K), continuous, gaseous ionized medium.
 Fluctuations in this intergalactic medium
(IGM) tend to have overdensities within a
factor of 10 of the cosmic mean and resemble
morphologically the filaments, walls and voids seen on larger scales
in the galaxy distribution at lower redshifts.
The dominant physical processes responsible for the state of this
IGM and the \lya\ absorption produced by it were anticipated by
 semi-analytic modelling of the \lya\ forest 
(e.g., McGill 1990, Bi, Borner \& Chu 1992). For completeness, 
we will summarize these processes below.

\subsection{The Fluctuating Gunn-Peterson Approximation}

 The  physical state of most of the volume of the baryonic
IGM  is governed  by
 the photoionization heating
of the UV radiation background, and the adiabatic cooling caused by
the expansion of the Universe. The competition between these two processes
drives gas elements towards a tight relation between temperature and
density, so that
\begin{equation}
T=T_{o}\rho_{b}^{\alpha}({\bf x}),
\label{rhot}
\end{equation} 
 where $\rho_{b}({\bf x})$ is the density of baryonic gas
 in units of the cosmic mean.
This relation holds well in simulations for $\rho_{b} \simlt 10$
(see e.g., Katz, Weinberg \& Hernquist 1996).
Hui \& Gnedin (1997) have explored the  relation  semi-analytically
by considering the evolution of individual gas elements in the Zel'dovich
Approximation. They find that the value of the parameters in equation
 (\ref{rhot}) depend on the history of reionization and
the spectral shape of the radiation background,
 and should lie in the narrow range  
$4000\;\K \simlt T_0 \simlt 15,000\;\K$ and  $0.3 \simlt \alpha \simlt 0.6$.

The optical depth for \lya\ absorption, $\tau$ is proportional to
the density of neutral hydrogen (Gunn \& Peterson 1965). In our case, this
is equal to the gas density $\rho_{b}$ multiplied by a recombination rate
 which is proportional to $\rho_{b}T^{-0.7}$. By using equation
(\ref{rhot}), we find that the optical depth is a power law function of the
local density:
\begin{equation}
\tau(x)=A\rho_{b}(x)^{\beta},
\label{tau}
\end{equation}
where $x$ is a distance along one axis, taken to the line-of-sight
towards the QSO (we are working in 
real-space for the moment). Because this result is simply a generalisation
of Gunn-Peterson absorption for a non-uniform medium, it has been dubbed
the Fluctuating Gunn-Peterson Approximation (FGPA, see Rauch \etal 1997,
Croft \etal 1998a, Weinberg \etal 1998a). The FGPA amplitude, $A$, is dependent
on cosmology and the state of the gas so that (e.g., Croft \etal 1999),
\begin{eqnarray}
A & = & 0.835
\left(\frac{1+z}{4}\right)^6 
\left(\frac{\Omega_b h^2}{0.02}\right)^2
\left(\frac{T_0}{10^{4}\;{\rm K}}\right)^{-0.7} \;\times \nonumber \\
& & \left(\frac{h}{0.65}\right)^{-1}
\left(\frac{H(z)/H_0}{4.46}\right)^{-1} 
\left(\frac{\Gamma}{10^{-12}\;{\rm s}^{-1}}\right)^{-1}\; .
\label{afacs}
\end{eqnarray}
Here $\Gamma$ is the 
photoionization rate, $h=H_{0}/100 \kmsmpc$, and
$\Omega_{b}$ is the ratio of the baryon density to the critical density.
The FGPA slope, $\beta = 2-0.7 \alpha \simeq 1.6$.  

The FGPA has been tested in simulations (Croft \etal 1997,
Weinberg 1999), and the predicted tight
correlation found to hold well.
 The analysis in this paper will involve  using
equation (\ref{tau}) to relate the optical depth to the
underlying real-space mass density. We will make predictions
for the observable quantity, transmitted flux in a QSO spectrum,
which we label $\phi$:
\bea
\phi(x) &=& e^{-A \rho_{b}(x)^\beta}.
\label{flux}
\eea
Equation (\ref{flux}) can be thought of as ``local biasing relation''
between the flux and mass distributions.
It can be seen that in this relation, the only spatially varying quantity
is $\rho_{b}(x)$ (ignoring global redshift evolution
and assuming a smooth ionizing background). Given that the
physical processes included in the derivation of the FGPA relation
 are the dominant ones, then the clustering properties of the \lya\ forest 
should be determined mainly by the statistics of $\rho_{b}$. 
The emphasis in this paper is therefore on applying our knowledge of the
behaviour of density perturbations to the \lya\ forest. We will use 
 analytical results for the non-linear evolution of
density perturbations 
in an effort to understand the origin of the values
of \lya\ forest observables. The ultimate aim is that 
with this understanding, measurements made from observational 
data can be used to directly
test both the gravitional instability hypothesis,
and the picture of the \lya\ forest outlined above,
as well as throwing light on the nature of the primordial density 
fluctuations. 

 An alternative to the  approach we adopt here would be
to use the local relation Eq[\ref{flux}] to directly
reconstruct  the density field, rather than to predict its cumulants
or the cumulants of the flux.
This reconstructed field could then be used 
to estimate the statistical properties of the density  (e.g.,  cumulants)
in a straightforward way.
This is not however simple to do in practice because
of the saturation of flux in high density regions.
Although large changes in high density regions 
have little effect on the statistics of the flux (i.e. the cumulants),
they will totally change the statistics of the density.
Any reconstruction technique will therefore have to deal with this missing
 information somehow. One approach for dealing with this
problem has been presented by Nusser \& Haehnelt (1998). 
In the present paper, we make use of the important
fact that the power-law
and exponential weighting of the density in the FGPA relation
results in a flux distribution whose statistical
properties are dominated by the small fluctuations, i.e. the linear or weakly
non-linear regime.
 
\subsection{Additional complications}

There are a number of assumptions concerning the relationsip
between flux and mass which we should discuss before
proceeding. First, the above equations  apply to the density of gas,
$\rho_{b}$ rather than the total density of matter, $\rho$, which
will be dominated by a dark matter component in the models we are considering.
At the relatively low densities of interest here, the distribution of gas in 
simulations does however trace the dark matter well.
 Pressure forces on gas elements tend to be small compared to the
gravitational forces, and non-hydrodynamical N-body simulations can be used 
to produce very similar spectra to the simulations which include these pressure
effects (Weinberg \etal 1999). Simulations do have finite resolution 
limitations, though, and clustering
in  a dissipationless dark matter distribution with
 power extending to small scales cannot be followed with infinite resolution.
The N-body only calculations so far used (e.g., Croft \etal 1998a)
have a resolution comparable to the small scale smoothing produced by pressure 
effects. Hydrodynamical simulations at high resolution (e.g., Bryan
 \etal 1998) can be used to study this smoothing. In the case of 
analytic work, one can first consider the linear regime.
In this case, the power spectrum
of fluctuations in the gas density, $P_{g}(k)$ is a smoothed version
of the dark matter power spectrum, $P_{\bf DM}(k)$, so that 
\begin{equation}
P_{g}(k)=\frac{P_{\bf DM}(k)}{[1+(k/k_{j})^{2}]^{2}}
\end{equation} 
where $k_{j}$ is the Jean's wavenumber (see e.g.,
Peebles 1993). In tests of this result,
Gnedin \& Hui (1998) have shown that after reionization, the effective
 smoothing length is generally smaller, and modelling with a different 
(Gaussian) filter
tends to give better results when compared with simulations. 
The situation in the non-linear regime will be more complicated.
The Jeans 
length scales as $(\rho_{b})^{-0.5}$,  but
 due to the temperature density relation
of equation \ref{rhot}, denser regions also tend to have higher 
temperatures, more thermal pressure, and more smoothing, 
so that the overall density dependence of the Jean's length should 
be weak.  Gnedin \& Hui (1998) show that filtering the initial 
conditions of a dissipationless simulation
with a single scale gives reasonable results compared to 
the full hydrodynamic case (although worse than their ``Hydro-PM''
technique, which involves adding a pressure term to the dissipationless
simulation calculations).
 In our case, the analytic approximation for
gravitational collapse which we use allows for filtering 
the evolved density with a top-hat 
filter in real-space (see the next sections and Appendix A1.2).
It may be possible to vary the smoothing length as a function of
density, but for reasons of simplicity we use a constant smoothing radius
for now. Another possibility for the future might be 
self-consistent modelling of the 
hydrodynamic effects when following the evolution of 
density perturbations. This has been done numerically in 1D simulations
 of spherical collapse by Haiman \etal (1996).

Second, the FGPA itself will break down in regions of
high density, because of shock heating of gas, collisional ionization,
star formation, and other processes.
We can quantify this by appealing to the results of hydrodynamic
simulations. As stated above, the relation has been directly tested by 
Croft \etal (1997), who find that it works well at high redshifts, $z\simgt 2$,
on a point by point basis, for $\rho_{b} \simlt 10$.
 When we consider statistics that we might want to measure from the flux 
distribution, the situation is even better. For example, we can see using the
 numbers given for $A$ above at $z=3$ and equation (\ref{flux}),
that optical depth will saturate ($\phi \simlt 0.05$)
for $\rho \simgt 3$. The physical processes occuring in  regions 
with $\rho \simgt3$
 are therefore not likely to  directly affect what we can measure.
Of course, there will be indirect effects, for example, supernova
winds may inhomogeneeously heat the IGM out into lower density regions.
Also,  the reionization of HeII, which is expected to occur around 
$z \sim 3$, may cause inhomogeneous heating if it is patchy enough 
(Miralda-Escud\'{e} and Rees 1994).
Although we expect the volume occupied by regions which do follow the FGPA
to be overwhelming in the high-z Universe, the statistical properties
of the absorption predicted by analytical gravitational instability theory 
should be useful in testing the validity of this assumption.  
They should also help us decide if their is any appreciable
contribution to clustering from spatial variations in the photoionization
rate $\Gamma$, due to the inhomogeneity in the UV background. Any such 
variations are expected  to be small in amplitude and to occur
 only on scales larger than we can probe directly at present (see e.g.,
Zuo 1992, Fardal \& Shull 1993, Croft \etal 1999). 

Third, we have so far only dealt with the density field in real-space,
whereas measurements from QSO spectra are made in redshift-space. Both peculiar
velocities and thermal broadening of absorption features should affect
the statistics of $\phi$ to some degree. We will include both these effects in 
our predictions.

\section{The probability distribution}

We would like to make predictions for the one-point PDF
of the flux $\phi$ and its moments.
The one-point PDF of a given field $\phi$ is defined so that 
the probability of finding, at a random position $x$, a value $\phi(x)$
 in the infinitesimal interval $\phi$ to $\phi+d\phi$, is  $P(\phi)d\phi$ 

To make these predictions we will first derive the corresponding 
probabilities $P(\rho)$, for the local matter overdensity $\rho=1+\delta$,
where $\rho$ is in units of the mean density $\rho(x)=n(x)/\langle n\rangle$.  
We will make indiscriminate use of either $\de(x)$ or $\rho(x)$ as 
variables when describing density fluctuations. The second
step is the assumption of a local relation with the form $\phi=f(\rho)$, 
(motivated by equation [\ref{flux}]).
 The PDF of the flux
will then simply be obtained by performing a change of variables
from the PDF of the density.

We start by assuming  a (Gaussian) form for the PDF of the initial 
conditions, and then follow its
 evolution . As we will see, in our approach
it is not necessary to assume Gaussian initial conditions, 
and this 
procedure can be  
extended to some other non-Gaussian models. 
We will do this with a model starting from $\chi^{2}$ initial 
conditions in Section 6.1.

One important point to note about our predictions is that we are not creating
artificial spectra but instead using an analytical model to evolve the
density PDF and then predict the PDF of the flux directly.
 Our predictions for the
density distribution will depend on only two parameters, equivalent to the
slope and amplitude of the linear correlation function on 
the smoothing scale, $\gamma$, and $\sigma^{2}_{L}$
 (see Sections 3, 4 and Appendix A1). This will allow us to cover
 a wide range of possiblities, and make predictions that are as generic
as possible.

In this section we will also test the effects of varying the two parameters
in the FGPA relation, $A$ and $\beta$, which (equation [\ref{afacs}])
contain information about
the cosmic baryon density, ionizing background and reionization history of
 the Universe. 

\subsection{The PDF of the initial conditions}

In the limit of early times, we assume an nearly homogeneous distribution with
very small fluctuations (or seeds), with given statistical properties.
We will concentrate on the case where the statistics of the initial density are
well described by Gaussian initial conditions, which correspond to a
general class of  models for the initial conditions. 
The one-point Gaussian probability distribution of 
an initial field $\delta$ is given by:
\beq
P_{IC}(\delta)~ = ~{1\over{\sqrt{2\pi \sigma^2_0}}}~\exp{\left(-{1\over{2}}
\left[{\delta\over{\sigma_0}}\right]^2\right)}
\label{p1gic}
\eeq

As the overdensity must be positive, $\rho>0$, we have that $\delta >-1$,
and a Gaussian PDF only makes physical sense 
 when the initial variance is small: $\sigma_0 \rightarrow 0$.

\subsection{The evolved mass PDF}

Because of  gravitational growth, 
the  evolution of $\delta$ 
will change the PDF from its initial form.
For small fluctuations linear theory provides a simple
way of predicting the time evolution  
of $\delta(t,x)= D(t) \delta_0(x)$, where $D(t)$ is the growth factor
(equal to the scale factor $D=a$ for $\Omega=1$),
and $\delta_0(x)$ is  the initial field. We will denote
 this linear prediction by $\delta_L$. For Gaussian initial conditions
the linear PDF is
also Gaussian with a variance $\sigma^2_L$, given by scaling the initial
variance $\sigma^2_0$ by $D^2$, so that  $\sigma^2_L= D^2 \sigma^2_0$.

As mentioned in the introduction,
there are a number of studies which predict the evolution of the PDF
beyond linear theory.
Here we will consider a generic class of
local mappings along the lines introduced by Protogeros \& Scherrer (1997).
The idea is for us to relate the non-linear fluctuation $\delta(q)$
(in Lagrangian space) with the corresponding 
linear fluctuation $\delta_L(q) \equiv D \delta_0(q)$
using a universal (local) function. To simplify notation we choose
to express this mapping as a relation between the
non-linear overdensity $\rho=\delta+1$
and the linear fluctuation $\delta_L$, so that 

\beq
\rho(q) = \calG[\delta_L(q)].
\label{eq:mapping}
\eeq

One such mapping is the spherical collapse model (SC)
or shear-free approximation.
For Gaussian IC, the SC approximation happens to 
give the exact statistical properties of the density
 (cumulants of arbitrary order)
at leading order in  perturbation theory 
(as found by Bernardeau 1992 in the context of
the cumulant generating function),
and provides a very good approximation to higher orders
(see FG98). Physically, 
this mapping corresponds to taking
the limit where shear is neglected. In this case the
equations for the growth of $\delta$, in Lagrangian space, 
are identical to those given by spherical collapse. So, in the
perturbative regime, the SC is the best mapping possible, given the
local assumption made in equation (\ref{eq:mapping}). The local
transformation naturally occurs in Lagrangian space
$q$ (comoving with the initial fluid element). 
The important point to notice
here is that although the local mapping is not the exact
solution to the evolution of $\delta$ (which is in general
non-local), it does give the correct clustering
properties in the weakly non-linear regime.
In the Appendix we give some specific examples for
the transformation $\calG$.

The one-point PDF induced by the above transformation,
in terms of the initial one-point PDF $P_{IC}$, is
\beq
P_L(\rho) =
P_{IC}(\de_L) \left|{d\de_L\over{d\rho}}\right|,
\eeq
where $\de_L=\calG^{-1}[\rho]$. As mentioned before, 
the above expression corresponds to the probability distribution
of the evolved field in Lagrangian space, $q$.
To relate Lagrangian and Eulerian probabilities we
use the law of mass conservation: $d\de(q)= \rho~ d\de(x)$, where
$\rho(x)=1+\de(x)$ is the overdensity in Eulerian coordinates.
We therefore have
\beq
P(\rho) = {1\over{N}}
{P_{IC}(\de_L)\over{\rho}} \left|{d\de_L\over{d\rho}}\right|,
\label{nlpdf}
\eeq
where $N$ is a normalization constant.
We will show some of these predictions in Section \ref{sec:sim} (e.g.,
Fig. \ref{pdf2}).

\subsection{The PDF of the flux}
\label{sec:bias}

We next assume a  local transformation which relates the
underlying smoothed overdensity to some observable 
quantity $\phi$:

\beq
\phi = f(\rho) 
\label{bk}
\eeq
This quantity can further be related to the linear density field, so that

\beq
\phi = f[\calG(\delta_L)].
\eeq

The PDF of $\phi$ will then related to that of the 
density by a simple change of variable:
\beq
P(\phi) = P(\rho) \left|{d\rho\over{d\phi}}\right| =
{1\over{N}} {P_{IC}(\de_L)\over{\rho}} 
\left|{d\de_L\over{d\phi}}\right|,
\label{fluxpdf}
\eeq
where $\de_L=\calG^{-1}[f^{-1}[\phi]]$
and $\rho=f^{-1}[\phi]$. Thus, given the transformations
$f$ and $\calG$, the above equations provide us with analytical (or
maybe numerical) expressions for the PDF of $\phi$.

In the present work, we concentrate on the cumulants
(see Section \ref{sec:pdf} for a definition) of the flux, rather
than the PDF itself. The reason for this is that,
 given the local assumption, the cumulants are more accurately determined
(see  Section \ref{sec:pdf} for more details). Nevertheless, we will
see in Section \ref{sec:sim} that the above prediction gives a good
 qualitative description of the PDF (see e.g., 
Fig. \ref{fpdf}).

\subsection{Redshift-space distortions}

The smoothed flux and its corresponding  optical depth $\tau=-\ln{\phi}$ 
has been  assumed
to be a local function of the {\it smoothed} non-linear
density $\rho$. The optical depth $\tau$ 
at a given real-space position along
the line of sight $r$  will  lie at a redshift-space 
position $s$ in a QSO spectrum:

\beq
s = r + v_r/H,
\eeq
where $v_r$ is the component of the smoothed  peculiar velocity 
along the line of sight  at $r$. Note that the
redshift distortion is of the smoothed field, where the smoothing,
due to finite gas pressure, occurs in real-space.
The redshift mapping will conserve optical depth,
$\tau_s ds = \tau dr$,  so that we have:
\beq
\tau_s = \, \tau \, \left|{dr\over{ds}}\right| = \, \tau \, \left| 1+ {1\over{H}} {dv_r\over{dr}} \right|^{-1}.
\eeq
In general, the relation between $dv_r/dr$  and $\rho$ will be complex.
However, in the SC model, spherical symmetry leads to 
 a great simplification:
\beq
{dv_r\over{dr}} = {1\over{3}} \nabla \cdot v \equiv \, {H\over{3}} \, \theta ,
\eeq
as, by symmetry, derivatives are the same 
in all directions (this idea has also been used
by FG98 and by Scherrer \& Gazta\~{n}aga 1998). 
We can now again use the local mapping to relate 
velocity divergence to the linear field: 
$\theta=\calG_v[\delta_L]$, as in equation (\ref{theta}).
Thus we have that the redshift optical depth is
given by a different mapping:
\beq
\tau_s(\rho) = \tau(\rho) \, \left|1+ {1\over{3}} \, \theta(\rho)\right|^{-1}.
\label{tauz}
\eeq
The redshift-space flux is simply:
\beq
\phi_s= exp[-\tau_s(\rho)],
\label{phiz}
\eeq
and its PDF can be computed with a simple change of variables:
\beq
P(\phi_s) = {1\over{N}}
{P_{IC}(\de_L)\over{\rho}} 
\left|{d\de_L\over{d\phi_s}}\right|.
\label{fluxpdfz}
\eeq

\section{The cumulants}

\subsection{Definitions}

Consider a generic field, $\phi$, which could be either
the measured flux in a 1D spectrum, $\phi=\phi(\rho)$, or the mass density in 3D 
space $\phi=\rho$. The $J$th-order (one-point) moments of this field are 
defined by (note the subscript ``c'' for connected):

\beq
m_{J} = \langle\phi^J\rangle = \int P(\phi) ~  \phi^J ~ d\phi.
\label{mij}
\eeq
Given the above relations we can choose to calculate  the moments
by integrating over $\delta_L$:

\beq
m_J = \int d\delta_L \, {P_{IC}[\delta_L]\over{\rho(\delta_L)}} \,
   \phi^J(\rho(\delta_L)),          
\eeq
or over the non-linear overdensity $\rho$:
\beq
m_J = \int d\rho \, {P_{IC}[\delta_L(\rho)]\over{\rho}} \,
\left|{d\delta_L\over{d\rho}}\right| \,   \phi^J(\rho).          
\eeq
Here $\phi$ can refer either to real-space fields or fields
which have been distorted into redshift-space 
(e.g., equation [\ref{phiz}]).
The $J$th order {\it reduced} one-point moments, or
cumulants  $k_J$, of the field $\phi$ are defined by:

\beq
k_{J} \equiv \langle\phi^J \rangle_c = \left. {{\partial{ \log M(t)}}
\over{\partial t}}\right|_{t \rightarrow 0},
\label{kij}
\eeq
where $M(t)=\langle\exp(\phi t)\rangle$ 
is the generating functional of the 
(unreduced) moments:
\beq
m_{J} \equiv \langle\phi^J \rangle = \left. {{\partial{M(t)}}
\over{\partial t}}\right|_{t \rightarrow 0}.
\eeq

The first reduced moments are:
\bea
k_{1} &=& m_{1}  \\
k_{2} &=& m_{2} - m_{1}^2 \nn \\
k_{3} &=& m_{3} -3 m_{1} m_{2}+ 2 m_{1}^3  \nn \\
k_{4} &=& m_{4}- 6 m_{1}^4 +12 m_{1}^2m_{2}  - 3 m_{2}^2 -4 m_{1}m_{3} \nn
\eea
and so on. Note that even when we normalise the flux so that the
mean is zero ($m_1=0$), the cumulants are different 
from the central moments in that the lower order
moments have been subtracted from them, so that  $k_{4} = m_{4} -3  m_{2}^2$
for $m_{1}=0$. 

It is interesting to define the following one-point hierarchical
constants:
\beq
S_J = {k_{J}\over{k_{2}^{J-1}}} ~~~~~~~~ J>2
\label{sj}
\eeq
This quantities turn out to be roughtly constant under
gravitational evolution from Gaussian initial conditions
(see e.g., Gazta\~{n}aga \& Baugh 1995 and references therein).

\subsection{Fully non-linear predictions}

We now use the one-point flux PDF obtained from equation (\ref{fluxpdf}) 
to predict the one-point
moments of the flux.
As mentioned in Section 3, 
 to make the predictions we need
the value of the linear variance $\sigma^2_L$, and 
the local relation equation (\ref{eq:mapping}), which only
depends on the smoothing slope $\gamma$
(see equation \ref{slopedef} for the definition of $\gamma$).
 For non-linear mapping
relations, we try each of the two cases
introduced in the Appendix. In the following figures we will 
only plot results for the Spherical Collapse (SC) mapping  because
they coincide  perfectly with the results for the Generalized 
Zel'dovich Approximation (GZA) model in equation (\ref{GZA})
with $\alpha=21/13$. 

We are interested primarily in the evolution of the density PDF
and the weighting which the FGPA relation gives to 
the clustering properties of the density field.
In order to separate the effects of redshift distortions from those 
of density evolution, we will present results in real-space
first. 

\subsubsection{The mean flux}

\begin{figure} 
\centering
\centerline{\epsfysize=8.truecm 
\epsfbox{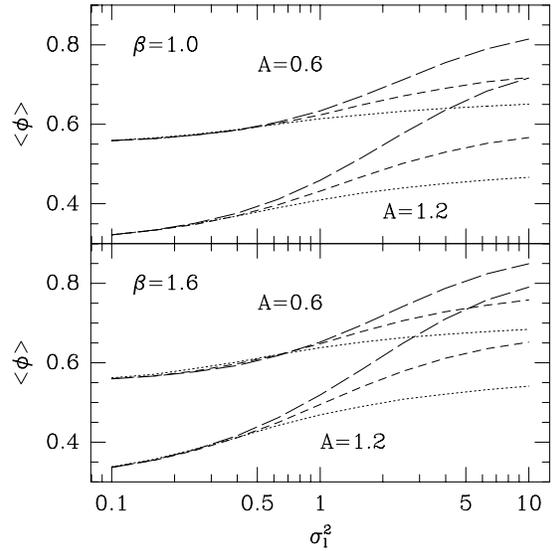}}
\caption[junk]{Mean flux $\langle\phi\rangle$ as a function of the linear
variance $\sigma^2_L$ for different values of $A$
and $\beta$. The dotted,  short-dashed and long-dashed
lines show the predictions for $\gamma=0,-1$ and $-2$
respectively.
}
\label{meanf}
\end{figure}

Figure \ref{meanf} shows the mean flux 
$\langle\phi\rangle$ ($=m_{1}$ in equation [\ref{mij}])
as a function of $\sigma^2_L$ for several values of
$A$, $\beta$ (the parameters in the FGPA relation) and $\gamma$.

For small $\sigma_L$ all predictions tend to 
$\langle\phi\rangle= exp(-A)$ which corresponds to the flux
at the mean overdensity: $\phi \rightarrow  exp(-A)$ 
as $\rho \rightarrow 1$. We will see in next subsection
that this is the leading order PT prediction.
For larger $\sigma_L$ the mean flux becomes larger, as 
expected, but it is  flatter  as a function
of  $\sigma^2_L$ when there is less smoothing, 
i.e., when $\gamma$ is less negative. Less smoothing of the density
will correspond to larger non-linearities, at least
in PT (see e.g., FG98). This seems to indicate that 
the effect of non-linearities is to {\it reduce} 
the mean flux, competing with 
 linear growth, which {\it increases} the mean flux.

\subsubsection{The variance of the flux}

\begin{figure} 
\centering
\centerline{\epsfysize=8.truecm 
\epsfbox{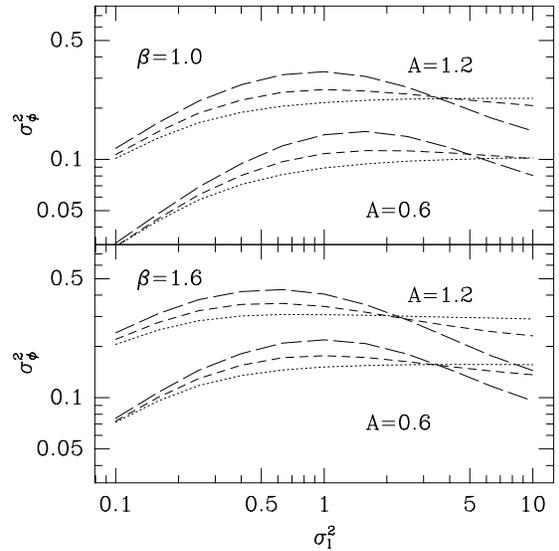}}
\caption[junk]{
 The variance of the
flux $\sigma_\phi^2$ as a function of $\sigma^{2}_{L}$.
The meaning of the line types is the same as in Figure \ref{meanf}.
}
\label{varf}
\end{figure}

We define the variance using the normalized flux:
\beq
\sigma^2_\phi \,=\, \langle\left({\phi-\langle\phi\rangle\over{\langle\phi\rangle}}\right)^2\rangle_c.
\label{eq:varf}
\eeq
The overall normalization by $\langle\phi\rangle$ is a convention,
in analogy to what is done for density fluctuations.
Figure \ref{varf} shows the predicted variance 
$\sigma^2_\phi$ as a function of $\sigma^2_L$ 
for several values of $A$, $\beta$ and $\gamma$.
For small  $\sigma^2_L$  these results reproduce
the linear relation: $\sigma^2_\phi= b_1^2 \,\sigma^2_L $
(see equation [\ref{eq:ptflux}]). Deviation from this power-law relation
(of index 1) occurs as $\sigma^{2}_{L}$ is
increased, and occurs sooner for larger values of $A$ and
$\beta$. For $\gamma >0$, when  $\sigma^2_L$ reaches $\sim 1$, 
$\sigma^2_\phi$ seems to reach a maximum
and then decreases again like a power-law for large  $\sigma^2_L$.

We can again see  that the predictions become flatter as a function
of  $\sigma^2_L$ when there is less smoothing, 
i.e., for less negative $\gamma$.

\subsubsection{The skewness of the flux}

\begin{figure} 
\centering
\centerline{\epsfysize=8.truecm 
\epsfbox{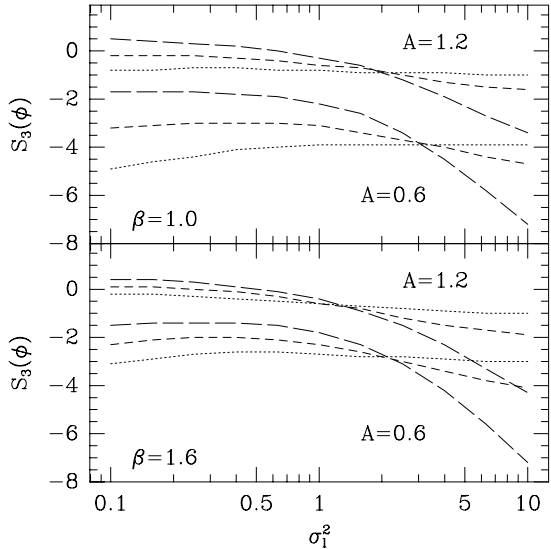}}
\caption[junk]{
 The Hierarchical skewness of the
flux, $S_3(\phi)$, as a function of $\sigma^{2}_{L}$.
The meaning of the line types is the same as in Figure \ref{meanf}.
}
\label{s3f}
\end{figure}

In a similar way, we define the (normalized hierarchical) 
skewness of the flux as:
\beq
S_3(\phi) \,=\, {\langle\left({\phi/{\langle\phi\rangle}}-1\right)^3\rangle_c\over{\sigma_\phi^4}}.
\label{eq:s3f}
\eeq
Figure \ref{s3f} shows the predicted skewness, as a function 
of $\sigma^2_L$ for several values of
$A$, $\beta$ and $\gamma$.
Because of the dependence of flux on density in the FGPA relation
(more density, less flux), while the density distribution
is positively skewed, the skewness of the flux tends to be negative
for most values of the parameters.

For small  $\sigma^2_L$  the results tend to a constant
as expected in leading order PT
(see equation [\ref{eq:ptflux}]).
We will examine the PT relations in more detail in Section 4.3.
For the moment, we note that
there again seems to be less variation in  $S_3({\phi})$ as a function
of  $\sigma^2_L$ for cases with less smoothing 
  (less negative $\gamma$).

\subsubsection{The kurtosis of the flux}

\begin{figure} 
\centering
\centerline{\epsfysize=8.truecm 
\epsfbox{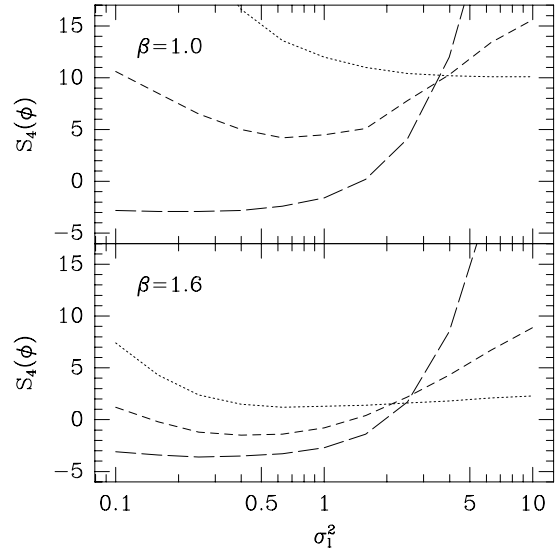}}
\caption[junk]{Same as Figure \ref{varf} for
the Hierarchical kurtosis. For clarity only $A=0.6$ is shown.}
\label{s4f}
\end{figure}

 The (normalized hierarchical) 
kurtosis of the flux is defined in a similar fashion:
\beq
S_4(\phi) \,=\, {\langle\left({\phi/{\langle\phi\rangle}}-1\right)^4\rangle_c\over{\sigma_\phi^6}}.
\eeq
Figure \ref{s4f} shows the predicted kurtosis, as a function 
of $\sigma^2_L$ for several values of
$\beta$ and $\gamma$. For clarity we only show one value of $A$.

For small  $\sigma^2_L$  these results tend to the constant value
predicted by leading order PT
(see equation [\ref{eq:ptflux}]). 
Being a fourth moment, $S_4{\phi}$ is extremely sensitive to deviations from
Gaussianity which occur as the density field evolves and
$\sigma^2_L$ increases. This sensitivity seems to be larger for lower
values of $\beta$, presumably because when $\beta$ is high, high density parts
of the PDF which might contribute heavily to the kurtosis of the
density field have little weight in the statistics in $\phi$. 

\subsection{Perturbative predictions}

An alternative to using the PDF is to calculate the cumulants directly
from the perturbative expansion along the lines suggested
(in the context of galaxy biasing) by Fry \& Gazta\~naga (1993).
That is we take:

\bea
\phi &=& e^{-A \rho^\beta}= b_1 \sum_{k=0} \, {c_k\over{k!}} \, \delta^k \\
b_1 &=& - A \, \beta  \,  e^{-A}\nn \\
c_0 &=& -{1\over{A\beta}} \nn \\
c_1 &=& 1 \nn \\
c_2 &=&  -1 +\beta -A\beta \nn \\
c_3 &=&  2-3 \beta+\beta^2 + 3 A\beta - 3 A\beta^2 +A^2\beta^2 \nn \\
\eea
and so on.
From this expansion one can simply estimate the moments 
by taking mean values to the powers of $\phi$. 
The leading order terms in   $\sigma^2_L$ are:
\bea
\langle\phi\rangle &=& e^{-A} \\
\sigma_\phi^2&=& b_1^2 \sigma^2_L \nn \\
S_3(\phi) &=& {S_3 + 3 c_2\over{b_1}} \nn \\
S_4(\phi) &=& {S_4 + 12 S_3 c_2 +12 c_2^2+4 c_3\over{b_1^2}}\nn , \\
\label{eq:ptflux}
\eea
where $S_3$ and $S_4$ are the leading order
(hierarchical) skewness and kurtosis for the density field.
For Gaussian initial conditions they are: 
 $S_3=34/7+\gamma$ and $S_4=60712/1323+62/3\gamma
+7/3\gamma^2$ (both in the SC model and PT theory).
For non-Gaussian initial conditions, one would have to
add the initial contribution, e.g.,  $S_3= S_3^0+34/7+\gamma$.

\begin{figure} 
\centering
\centerline{\epsfysize=8.truecm 
\epsfbox{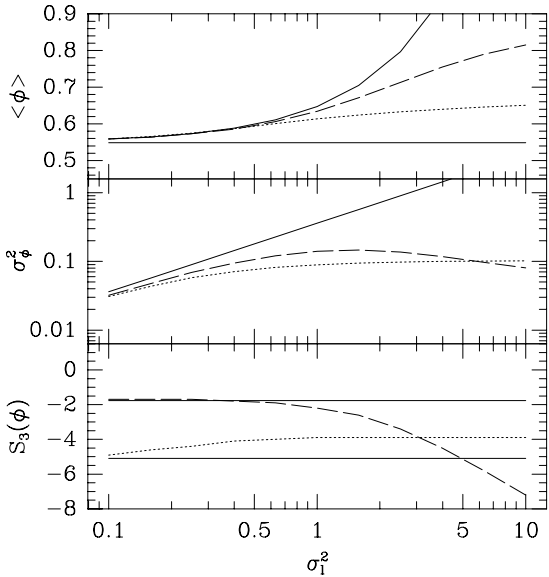}}
\caption[junk]{
Perturbative predictions for the one-point flux moments compared to the
fully non-linear predictions. 
Mean flux $\langle\phi\rangle$ (top), variance $\sigma^2_\phi$
(middle panel) and skewness $S_3(\phi)$ (bottom) 
are shown as a function of the linear
variance $\sigma^2_L$ for $A=0.6$
and $\beta=1.0$. The dotted  and long-dashed
lines show the non-linear predictions 
(Section 4.2) for $\gamma=0$ and $\gamma=-2$ smoothing
respectively. The straight continuous lines show the
corresponding  leading order perturbative prediction (Section 4.3),
valid for $\sigma_L \rightarrow 0$.
The solid curve in the top panel is the perturbative prediction
 for $\langle\phi\rangle$
including the effect of a higher order (loop) correction (see
equation [\ref{meanpt}]).
}
\label{ptflux}
\end{figure} 

These predictions are shown as a straight continuous lines
in Figure \ref{ptflux}, where they are compared to the
full (non-perturbative) calculation in the SC model for two
values of the smoothing slope, $\gamma=0$ (long-dashed line)
and $\gamma=-2$ (dotted line).
We can see that the expressions only converge on the
 correct result asymptotically as $\sigma_L \rightarrow 0$. 
The relative performance depends on $\gamma$, with steeper slopes giving
better results. 
It is easy to calculate higher order (loop) corrections
(see e.g., FG98). For example:
\beq
\langle\phi\rangle = e^{-a} \left[1+{A\beta\over{2}}(1-\beta+A\beta)\, \sigma^2_L + 
\Or(\sigma^4_L) \right],
\label{meanpt}
\eeq
This prediction for the mean flux is shown as a curved continuous line
in the top panel of Figure \ref{ptflux}, and seem to work up to scales where 
$\sigma_L \simeq 1$.
We find however that the agreement becomes worse for larger values 
of $A$ and $\beta$. A similar tendency is found for other moments. 

We have seen that even when high-order corrections are included,
this perturbative approach only works well for small values 
of $\sigma_L$, $A$ and $\beta$. It is likely to have only a limited
applicablity when we consider the situation in the observed Universe,
where typical values of $\sigma_L \geq 1$ are expected, (at
least for redshifts $z \simlt 4$).
Given that we have the  possibility of implementing the SC model (or 
the GZA model) to arbitrary order, one could ask why we bother
with a perturbative approach. 
The first obvious reason is that it gives us
 compact analytical predictions, simple formulae which are
 functions of the input variables ($A$, $\beta$, $\gamma$ 
and $\sigma^2_L$). A second reason is that by using this
approach, it may be possible to introduce the PT solutions.
As mentioned before, PT
only differs from the SC model through the shear contributions, and
although these are typically small (as will be shown) they might still
be relevant for higher precision comparisons. It is
not clear nevertheless than one could obtain higher
accuracy at $\sigma_L \simgt 1$,
given the limitations of a perturbative approach
(i.e., convergence of the series).

\subsection{Predictions in redshift-space}

\begin{figure} 
\centering
\centerline{\epsfysize=8.truecm 
\epsfbox{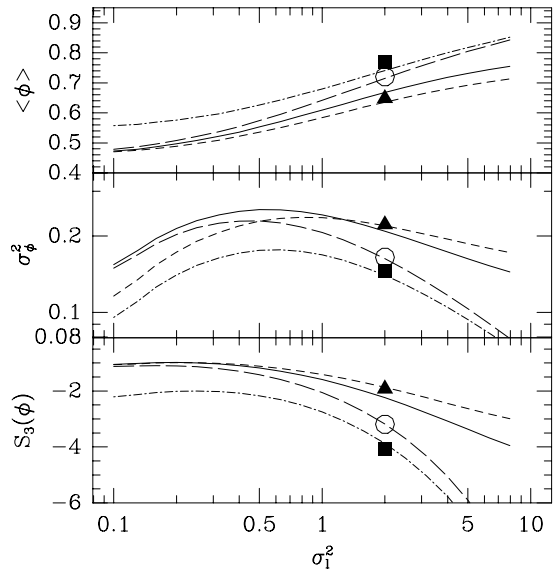}}
\caption[junk]{
The effect of redshift-space distortions on the one-point moments.
Mean flux $\langle\phi\rangle$ (top), variance $\sigma^2_\phi$
and skewness $S_3(\phi)$ (bottom) are plotted as a function of the linear
variance $\sigma^2_L$ for $\gamma=1$, $A=0.8$ 
and $\beta=1.6$. The short-dashed  and continous 
lines show the predictions in real and redshift-space 
(peculiar velocities only) respectively.
The long-dashed lines correspond to predictions in redshift-space with an
additional velocity dispersion component on small scales (added in
order to match simulation results - see text).  
The dot-dashed lines show the predictions in redshift-space with thermal
broadening as well as peculiar velocities (and the extra small-scale 
dispersion).
 In modelling the
thermal broadening component to the  redshift distortion
we have assumed that the temperature depends on the
 density, as predicted by equation (\ref{rhot})
(see Section 4.4).
The points show results from 
simulated spectra (set [a])
described in Section 5.
The triangles, circles and squares represent spectra in
real space, redshift-space with no thermal broadening,
and redshift-space with thermal-broadening, respectively.
}
\label{zflux}
\end{figure} 

We now turn to the more observationally realistic case
where redshift distortions are included.
Fig. \ref{zflux} shows how the (fully non-linear) 
predictions change when given in redshift-space.
We  use the formalism of Section 3.4 (e.g., equation [\ref{phiz}]),
results which allow one to estimate the effects of peculiar velocity
distortions. The redshift distortions caused by thermal broadening can 
be treated in a similar way, and we will also do this below.
For clarity we only show a single value of $A$, $\beta$
and $\gamma$, but similar effects are found if we use  
other values. 
We have also plotted some simulation points on Fig. \ref{zflux}.
The simulations will be decribed fully in the next section.
For the moment, it is only necessary to mention here that, for purposes
of comparison, the simulation
density PDFs can be decribed by the two parameters,
$\gamma$ and $\sigma^{2}_{L}$
(evaluated from the power spectrum of initial
fluctuations used to set up the simulation).
 The transformation from density into
flux in the simulations has also been carried out using the FGPA relation. We
plot points both with and without
 including  redshift distortions from peculiar 
motions and thermal broadening. 

If we concentrate on the mean flux first, and ignore thermal
broadening, we can see that 
$\langle\phi\rangle$ in redshift-space (continous line in the
top panel) seems to converge to  same value as the real-space mean flux
(short-dashed line) for small $\sigma_L$. For $\sigma_L \simgt 1$
the mean flux is larger in redshift-space.

The variance in the flux, $\sigma_\phi^2$, defined
by equation (\ref{eq:varf}) is shown in the middle panel of Fig.
 \ref{zflux}. We can see that this quantity is larger 
in redshift-space than in real-space for  $\sigma_L \simlt 1$, a trend that is
reversed for $\sigma_L \simgt 1$. The former is presumably due to the 
 same ``squashing effect'' that is seen in studies of
the density field, where infall of matter into high
density regions enhances clustering (Kaiser 1987).  The latter 
effect can be attributed to a relative decrease in the level of 
redshift-space clustering caused by high velocity dispersion along
the line of sight. 

The skewness, defined
by equation (\ref{eq:s3f}) is shown in the bottom panel of Fig.
\ref{zflux}. For small $\sigma_L$, the redshift-space 
(continous line) $S_3(\phi)$ seems to match the real-space value
(dashed line). For $\sigma_L \simgt 1$,
$S_3(\phi)$  is  smaller in redshift-space.

The simulation points on the plot in real space
(triangle) agree with predictions in the real space.
 In redshift-space (open circles), although the
sign of the change caused by the peculiar velocity distortions is correct,
the predictions do not agree in detail. Our interpretation of this
is that the SC model does not predict enough random non-linear 
velocity dispersion. We have therefore added a 
velocity dispersion by hand, by adding in a velocity
divergence term, $\theta_{\rm disp}$
 to equation (\ref{phiz}):
\begin{equation}
\tau_s(\rho) = \tau(\rho) \, \left|1+ {1\over{3}} \, [\theta(\rho)+
\theta_{\rm disp}(\sigma^{2}_L)] \right|^{-1}.
\label{tauztb}
\end{equation}
In order that the asymptotic behaviour of $\theta_{\rm disp}$ be satisfied
(non-linear dispersion $\rightarrow 0$ as $\sigma^2_L \rightarrow 0$),
we have adopted the functional form $\theta_{\rm disp}=C\sigma^2_L$, 
Predictions from the SC model including the effect of
this term are shown in fig. \ref{zflux} as a long dashed line.
We have adjusted the constant $C$ so that the predictions go through the 
simulation point in redshift-space.

Next, we include thermal broadening in our predictions.
For the moderate optical depths 
we are concerned with here, the relevant
Voigt profile can well be approximated by 
a Gaussian velocity dispersion. The width 
of this Gaussian profile,
 $\sigma_T \simeq \sigma_{T_0} (T/T0)^{1/2}$, where
$\sigma_{T0} \simeq 13/\sqrt{2} \kms$ for $T_0 \simeq 10^4 K$.
From equation (\ref{rhot}) we have $T \propto \rho^{0.6}$, so that 
$\sigma_T \simeq \sigma_{T0} \rho^{0.3}$. We can think of
 thermal broadening as resulting in the addition of 
a thermal velocity component, $\theta_T$, to the 
divergence field $\theta$ in equation (\ref{tauz}).
We will model this thermal component in a similar way to the extra
non-linear dispersion term ($\theta_{\rm disp}$)
defined above, which can be thought of
as a ``turbulent'' broadening term.
We simply model the additional thermal dispersion
 using its rms value, so that,

\beq
\theta_T(\rho) \simeq 3 \, {\sigma_T\over{H\Delta}}
\simeq {\sigma_{T0}\over{H\Delta}} \, \rho^{0.3},
\eeq
where  $\Delta$ is distance in the QSO spectrum
which corresponds to the scale of Jean's smoothing.
This density-dependent term then enters the RHS of equation (\ref{tauztb}),
alongside $\theta_{\rm disp}$.
The long-dashed line in Fig. \ref{zflux}
shows the effect of thermal broadening using this prescription.
We have used the value $\sigma_{T0}$
appropriate to the simulation (see Section 5.2 for details),
whose thermally broadened results are plotted as a solid square.
As can be seen, thermal broadening results in more flux being
absorbed, and yields a lower value of $\sigma_\phi^2$.
This is as we should expect, given that the distribution of optical
depth has effectively been smoothed out by the addition of a dispersion.

It is evident from these results that
 the one-point moments depend fairly
strongly on the details of redshift distortion modelling, which are likely
to be poorly constrained a priori in our approximate treatment.
 Fortunately, much of the uncertainty can be removed
by setting the mean flux, $\langle\phi\rangle$, to be equal to some 
(observed) value.
We will show later (Section 5.5) that by doing this, the moments can be made
insensitive to the inclusion of redshift distortions.

\section{Comparison to simulations}
\label{sec:sim}

As the emphasis of this paper is on 
the role of gravitational evolution of the density
field, we now test the analytic techniques 
we have employed against numerical simulations of gravitational clustering.
The  N-body only simulations which we use do  not allow
us to perform tests of the validity of the
 model we have assumed  for relating the mass distribution 
to an optical depth distribution
(the FGPA, Section 2). 
 Any difference between the statistical properties
of the \lya\ forest we predict and those observed in nature could therefore 
stem from a misapplication of the FGPA, rather than
 from problems with the underlying density field. 
 Tests performed in other contexts show that this is unlikely, 
as  a dissipationless
approach to simulating the \lya\ forest
 can perform well (see e.g. Croft \etal 1998a,
Weinberg \etal 1999) in comparison with the full hydrodynamic case.
Approximate methods should neverthless be tested on a case by case basis,
and we reserve comparisons with hydrodynamic simulations for future work.

\subsection{Simulations}

The simulations we use have all been run with a P$^{3}$M N-body code 
(Efstathiou \& Eastwood 1981, Efstathiou \etal 1985).
 The softening length of the PP interaction was made large (1 mesh cell),
for simulation sets (a-c) because high spatial resolution is not needed.
 We have not attempted 
to simulate any particular favoured cosmological models or even to make sure 
that cases are likely to be compatible with
 expectations for the nature of 
the density field at high redshift. We are more interested in spanning a wide
 range of values of $\sigma^{2}_{L}$, and $\gamma$, the parameters which 
determine
the nature of the evolved density field in the SC model.
To this end, we use outputs from three different sets of simulations. 
The initial conditions for all runs were Gaussian random fields
with CDM-type power spectra of the form specified by Efstathiou, Bond \& White
(1992), with a shape parameter $\Gamma$, so that
\begin{equation}
P(k) \propto \frac{k}{[1+(ak+(bk)^{3/2}+(ck)^{2})^{\nu}]^{2/\nu}}
\end{equation}
where $\nu=1.13$, $a=6.4/\Gamma \hmpc$, $b=3.0/\Gamma \hmpc$,
 $c=1.7/\Gamma \hmpc$. There are 
five realizations with different random
phases in each set of simulations, which are described below.

(a) A set with a box-size $40 \hmpc$ and shape parameter $\Gamma=0.5$.
The model was run to $z=3$ with an Einstein-de-Sitter cosmology.
At that redshift $\sigma^{2}_{L}=2.0$.
at the smoothing radius
(see below), which was $0.31 \hmpc$ (comoving). The linear slope
at the smoothing scale, $\gamma=-0.8$. 
 These simulations were run with $200^{3}$ particles and $256^3$ cells.

(b) A set with box size $22.22 \hmpc$, $128^{3}$ particles,
$\Gamma=0.5$, and an Einstein-de-Sitter cosmology.
 Simulated spectra were made from
this set assuming that $z=3$, but several different outputs were used with 
varying amplitudes of mass fluctuations.
The smoothing radius was $0.31 \hmpc$ (comoving),
on which scale the linear slope was $\gamma=-0.8$. The value of
$\sigma^{2}_{L}$ on this scale ranged from $0.02$ to $7.5$ for the different
outputs.

(c) A set the same as (b), except with
$\Gamma=10$.
The smoothing radius was again $0.31 \hmpc$ (comoving),
on which scale the linear slope was $\gamma=-1.8$. The value of
$\sigma^{2}_{L}$ on this scale ranged from $0.03$ to $10$ for the different
outputs.

(d) A set with box size $20 \hmpc$, $126^{3}$, particles and
$\Gamma=3.8$. This set 
was originally used as ensemble G in Baugh, Gazta\~{n}aga
\& Efstathiou 1995, but the box size was taken to be larger, and
hence $\Gamma$ was lower. The smoothing scale is $0.37 \hmpc$, where
  $\gamma\simeq -1.5$, and $\sigma^{2}_L=0.25$ (at $z=3$).
The cosmology assumed has $\Omega=0.2$ and $\Omega_{\Lambda}=0.8$
at $z=0$.

We have also run a single simulation with the same parameters as those in set
(a), except with a box of size $11.11 \hmpc$, and $128^{3}$ particles,
so that the mass resolution is increased by a factor of $\sim 12$.

\subsection{Simulated spectra}
To make simulated spectra from the N-body outputs, we use the following 
procedure (see also Hui, Gnedin \& Zhang 1997, Croft \etal 1998a).
The particle density and momentum
distribution is assigned to a ($256^{3}$) grid using a TSC (triangular
shaped clouds, Hockney \& Eastwood 1981) scheme.
The resulting fields are smoothed in Fourier space with a filter, which
in our case is a top-hat with radius given in Section 5.1 above.
We also use a very narrow Gaussian filter (with $\sigma$ 0.1 times 
the top-hat radius)
 in order to ensure that density is non-zero everywhere.  
The velocity fields are computed by dividing the momentum by the density 
everywhere. Spectra are selected as lines-of-sight through the box, parallel
to one of the axes (we select the axis randomly for each line-of sight).
These one-dimensional density fields are then converted to an optical depth 
distribution using equation (\ref{tau}). We also
compute  the temperature distribution of the gas
using equation (\ref{rhot}),
with $\alpha=0.6$ and $T_{0}=10^{4}$K. The optical depths are then 
converted from real-space to redshift-space by convolution with the
line-of-sight velocity field and with a Gaussian filter with the appropriate
thermal broadening width.

We estimate the one-point statistics of the flux in the spectra
without any additional smoothing using counts-in-cells
and the estimators of Section 4.2.
We extract 5000 spectra from each simulation realization
and estimate statistical errors from the scatter in results
between the 5 realizations. We find that the resulting
error bars are typically much smaller
than the symbol size and so we do not plot them in the figures.
Small systematic errors will arise because of the additional smoothing
involved in using a mass assignment scheme, also because
of shot noise from  particle discreteness. 

\subsection{The PDF of density and flux}
\label{sec:pdf}

\begin{figure} 
\centering
\centerline{\epsfysize=8.truecm 
\epsfbox{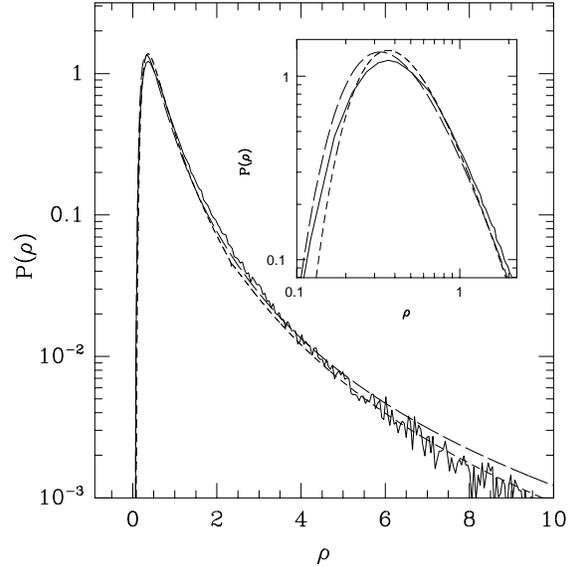}}
\caption[junk]{
The one-point PDF of the density field
$\rho$ measured from N-body simulations (continuous
line, set [a], see Section 5.1).
The simulation density field was
smoothed with top-hat cell on a scale with linear 
variance $\sigma_L^2 \simeq 2$ and slope $\gamma=-1$.
The prediction of two approximations to PT
are also shown: the Spherical Collapse model (short-dashed line)
and the GZA in $\alpha=21/13$ dimensions (long-dashed line).
}
\label{pdf2}
\end{figure} 

We first compare the PDF of the density in the simulations with
the analytical predictions, both in real space. 
The one-point density PDF estimated from simulation  lines-of-sight
(using simulation set [a])
is shown
in Fig. \ref{pdf2} as a continuous line. The analytical
 predictions are shown as a short-dashed (SC) and long-dashed (GZA) line.
In evaluating the predictions we have used $\gamma\simeq -1$ and
$\sigma_L^2 \simeq 2$ which correspond to the appropriate linear theory
values for simulation set (a). 

Note that the spherical collapse model is only an approximation to 
PT, so we do not expect to recover the PDF exactly, even 
close to $\delta \simeq 0$. 
To carry out the exact recovery we would have to include non-local (tidal) 
effects.  The mean tidal effects vary in proportion to the 
linear variance and the leading contribution (when variance goes to zero) 
is only local (but non-linear). Tidal effects seem to 
distort the PDF, turning some $\delta=0$ fluctuations into either voids 
 $\delta \simeq -\sigma$ or overdensities $\delta \simeq \sigma$, 
so that the real peak in the PDF 
is slightly lower than SC or GZA. 
This distortion is significant, given that the statistical errors 
on the simulation result are small.
 The overall shape is however similar,
as we can see from Fig. \ref{pdf2}. The lower order moments of the
density are 
also fairly similar: tidal effects tend to cancel out. 
When the variance is small (and the PDF tends to a 
Gaussian), the PDF can be defined one-to-one by its moments (e.g.,
with an Edgeworth
expansion). In this limit tidal effects vanish and both the PDF and
the moments are given exactly by the SC model.

Thus, for the density distribution, the statistics of 
the moments are dominated by the contribution of local dynamics to the PDF 
and tidal effects are subdominant (they tend to 
cancel out when taking the mean). 
It is plausible that a similar sort of cancellation will happen for the 
statistics of the flux.
In the next sections, we will therefore 
concentrate on predictions for the moments of the flux.

\begin{figure} 
\centering
\centerline{\epsfysize=8.truecm 
\epsfbox{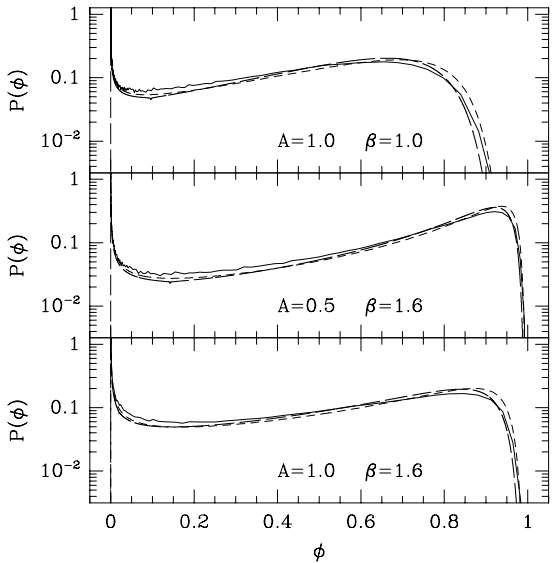}}
\caption[junk]{PDF of the flux evaluated using the same density distribution  
plotted in Fig. \ref{pdf2}. Three different values of
the FGPA parameters $A$ and $\beta$ are used as shown in each panel.
The one-point PDF in the simulations is shown as continuous lines and is
compared with  the predictions of the two PT approximations, 
the Spherical Collapse model (short-dashed line),
and the GZA in $\alpha=21/13$ dimensions (long-dashed line).
All results are in real-space.
}
\label{fpdf}
\end{figure}

Different density regimes will be prominent if we consider
the PDF of the flux, as it is a transformed version of the density PDF.
In Fig. \ref{fpdf}
 we apply the FGPA relation (ecuation 4) in real-space to the density PDF from 
Fig. \ref{pdf2}.
The resulting flux distributions are rather flat, with the high density
tail being confined to a small region of flux space near $\phi=0$.
Varying the parameters $A$ and $\beta$ produces most notable differences in
the fraction of spectra which show little absorption. We can see that
the lower $\beta$ is, 
the less likely it is that any pixels will be seen with
values near  the unabsorbed QSO flux. The specific case 
with $\beta=1.0$ is not realistic, though, as $\beta$ is expected to
be $\simgt1.6$ for all reasonable reionization histories
 (Hui \& Gnedin 1997).

\begin{figure} 
\centering
\centerline{\epsfysize=8.truecm 
\epsfbox{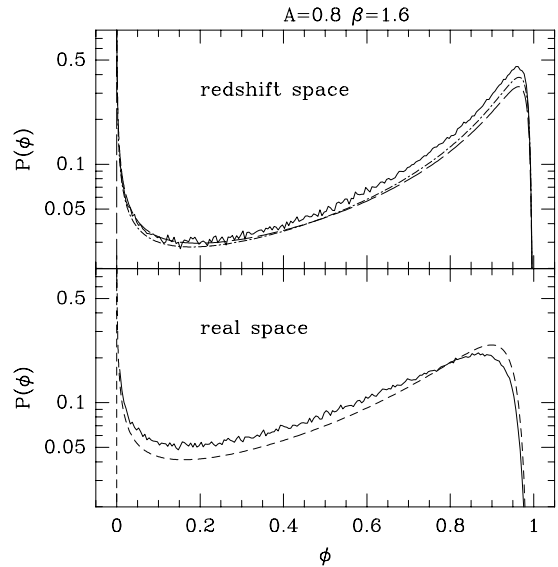}}
\caption[junk]{
PDF of the flux evaluated from the same density distribution  
used in Fig. \ref{pdf2} (simulation set [a]).
The one-point PDF in the simulations is
shown as a solid line. Real-space results are plotted in the 
 bottom panel, and results in redshift-space with thermal broadening
in the top panel.
These results are compared with the  appropriate analytical predictions,
using the Spherical Collapse model (short-dashed line)
and the GZA in $\alpha=21/13$ dimensions (long-dashed line).
}
\label{fpdfz}
\end{figure}

In Fig. \ref{fpdfz} we show the effect of redshift distortions on the  flux
PDF. We have used the prescription of Section 4.4 for making the 
predictions, including thermal broadening.
The effect of the distortions is to  evacuate the low density regions,
as we might expect. The effect is very noticeable, as the peak of the
PDF is raised substantially.
Note that to to make this plot, we have not added an additional
small scale velocity dispersion, as we did previously 
 to obtain
a closer match to the one-point moments
 in simulations (Fig. \ref{zflux}). 
We can see that the 
agreement of the predictions and simulations for the PDF
in Figure \ref{fpdfz} is still good,  showing that our redshift modelling
is at least a  good qualitative approximation
to the underlying physical processes.

\subsection{Cumulants of the flux}

\begin{figure} 
\centering
\centerline{\epsfysize=8.truecm 
\epsfbox{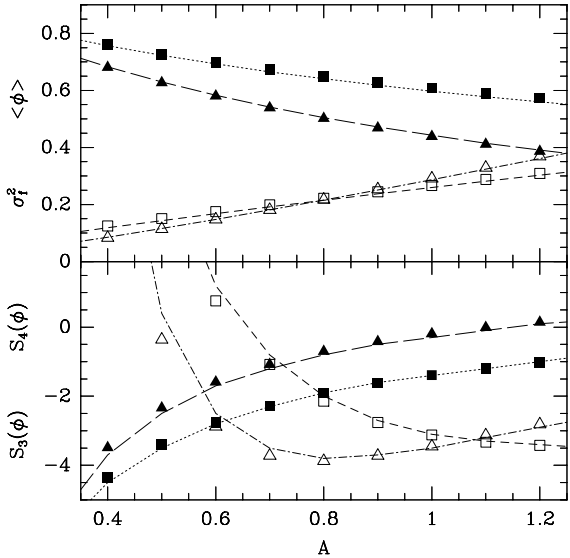}}
\caption[junk]{
The first 4 moments of the flux in  simulation  set (a)
(squares) and set (d) (triangles) compared to the appropriate
analytical predictions (lines),
as a function of $A$. The top panel
shows the mean flux $\langle\phi\rangle$ (closed figures)
and the variance $\sigma_f^2$ (open figures). 
The bottom panel shows the hierarchical skewness $S_3(\phi)$ 
(closed figures)  and the kurtosis $S_4(\phi)$ (open figures).
All results are in real-space.
}
\label{testac80}
\end{figure}

In Fig. \ref{testac80} we compare predictions and simulation
results (for simulation sets [a] and [d]) for the mean
flux $\langle\phi\rangle$, the variance $\sigma_\phi^2$, skewness
$S_3(\phi)$ and kurtosis $S_4(\phi)$. In  all cases we have
used $\beta=1.6$, and we show the moments as a function
of the  value of $A$.
All results in this fig. have been evaluated in real-space.
 Squares correspond to
simulation set (a), in which the one-point linear variance
of the density field, $\sigma_L^2=2$
(the non-linear value is $\sigma^2 \simeq 4$) 
and the linear slope on the smoothing scale is
 $\gamma=-1$. Triangles correspond
to  set (c), for which $\sigma_L^2=0.25$ and  $\gamma=-1.5$.
We can see that the predictions (from the SC model) 
are in good overall agreement with the simulations. It is
encouraging  that the agreement
is not noticeably worse for set (a), which has a much higher amplitude
of mass fluctuations on the relevant scale.

\begin{figure} 
\centering
\centerline{\epsfysize=8.truecm 
\epsfbox{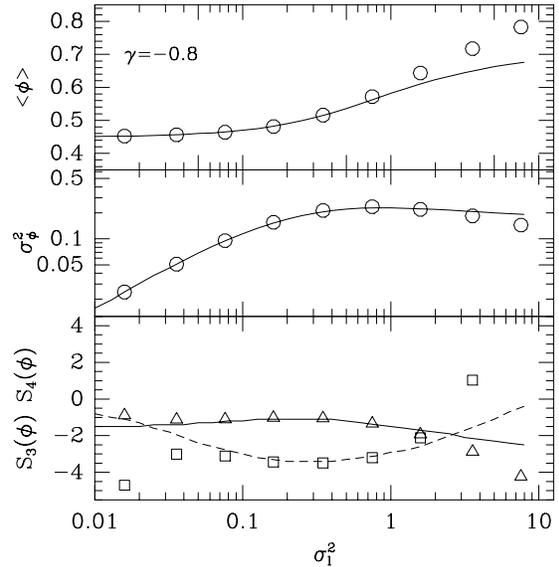}}
\caption[junk]{
The mean flux (top panel), variance (middle), $S_3$ (triangles), and 
$S_4$ (squares) for simulation set (b), as a function of $\sigma^{2}_{l}$.
We have used $A=0.8$ and $\beta=1.6$ to generate the spectra.
The corresponding predictions of the spherical collapse model are shown as
lines. All results are in real space.
}
\label{vssig1}
\end{figure}

\begin{figure} 
\centering
\centerline{\epsfysize=8.truecm 
\epsfbox{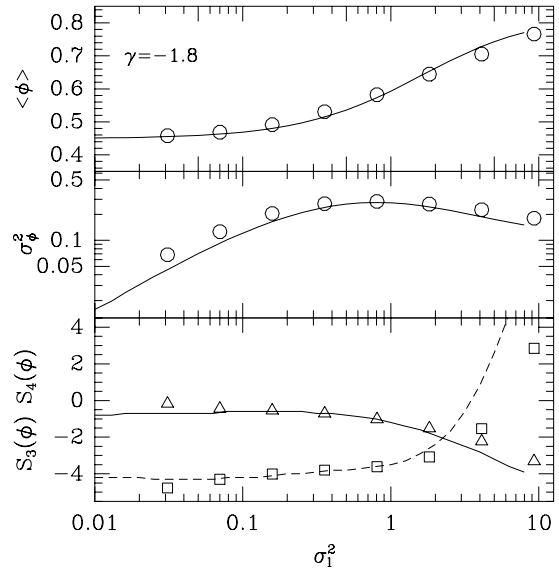}}
\caption[junk]{
The mean flux (top panel), variance (middle), $S_3$ (triangles), and 
$S_4$ (squares) for simulation set (c), as a function of $\sigma^{2}_{l}$.
We have used $A=0.8$ and $\beta=1.6$ to generate the spectra.
The corresponding predictions of the spherical collapse model are shown as
lines. All results are in real space.
}
\label{vssig2}
\end{figure}

We can also compare the predictions of the SC model with simulations
for different values of the amplitude of mass fluctuations. In Figures
\ref{vssig1} and \ref{vssig2} we have done this, by plotting the flux moments
for several different simulation output times. We can see that for most of
 the range, the predictions work well, for both sets of simulations with
 different linear slopes ($\gamma$).
At  very low amplitudes $\sigma^{2}_{L}$, the simulations will 
still be dominated by the effect of the initial particle grid, so that
the  differences we see for the lowest amplitude output are not surprising.
For high values of $\sigma^{2}_{L}$, the SC predictions start to break
down, with $S_3$ and $S_4$ being most affected. This shows that we must be
careful when intepreting results which appear to indicate high values
of $\sigma^{2}_{L}$. 

Ultimately, the real test of how far we should trust our 
analytic methods  should come from comparisons with 
hydrodynamic simulations, and in particular
those that been run at resolutions high enough to resolve the Jean's scale.
In such simulations (e.g., Bryan et al 1998, Theuns et al 1998) it is found 
that quantities such as the mean effective optical depth are indeed sensitive
to resolution. In the SC model computations,
although the smoothing scale does not enter directly as a parameter,
 the quantities which do enter, $\gamma$ and $\sigma^{2}_{L}$ are dependent
on it. A smaller Jean's scale will yield higher values of $\sigma^{2}_{L}$,
and as we can see from the N-body only tests of Figs. \ref{vssig1}
and \ref{vssig2}, the accuracy of our approximations can vary widely.
The first three moments of the flux are recovered within $\sim 10-20\%$, for
$\sigma^{2}_{L}\simeq 4$ and below (for which the non-linear variance in
the density field, $\sigma^{2} \sim 10$). The kurtosis of the flux has a much
 larger error, although as we will see later some non-Gaussian models
have such a different $S_{4}$, that it should still be detectable. 
A more direct test involves consideration of  results
from the higher resolution version of the simulations from set (a).
This simulation has a higher mass resolution by a factor of $\sim 12$.
When we use the same filter size as for set (a), we find that the flux 
moments do not change.
If we decrease the filter size by $12^{\frac{1}{3}}$ to $0.13 \hmpc$, 
(at which scale $\sigma^{2}_{L}=3.8$), the SC predictions for the mean flux
 and variance are still accurate to better than $10 \%$
, but the prediction for $S_{3}(\phi)$ is
$-2.3$, when the simulation value is $-3.0$, and the SC $S_{4}(\phi)$
is $-1.0$ when the simulation gives $+2.4$.

One way of looking at these statistics is as 
constraints on the unknown parameters
which describe the density field,  $\gamma$ and $\sigma^2_{L}$.
If we return to Fig. \ref{testac80} we
note that the results for $\sigma_\phi^2$ are quite
similar for both sets of simulations. This can be
also seen in Fig. \ref{varf}, if  we look
at the results for different
values $\beta$ and $A$.
The mean flux or the skewness
seem to respond more sensitively to  $\sigma_L$.
We can also see this by examining  Figures \ref{meanf} and \ref{s3f},
where for small $\sigma_L$ the skewness of the flux
is a much better indicator of $\gamma$ than
the mean flux. This then changes at  $\sigma_L^2 \simeq 2$
where there is a degeneracy in $S_3(\phi)$ for different
values of $\gamma$, while $\langle\phi\rangle$ seems to give different
predictions. 
It is interesting that, although one model has
a larger amplitude of mass fluctuations, the variance in the flux
is not systematically higher or lower in one model than the other, but 
instead which is higher depends on $A$.
We also note that the kurtosis becomes quite large
for small values of $A$. 
Although the accuracy of our predictions
for $S_{4}$ decreases as $\sigma_L^2$ becomes large (see above), when
$\sigma_L^2$ is moderate, the
rapid increase in $S_{4}$ seen for small values of $A$ is reproduced, 
 even for values as large as $S_{4}\simeq100$.

\subsection{Comparisons with the mean flux held fixed}

\begin{figure} 
\centering
\centerline{\epsfysize=8.truecm 
\epsfbox{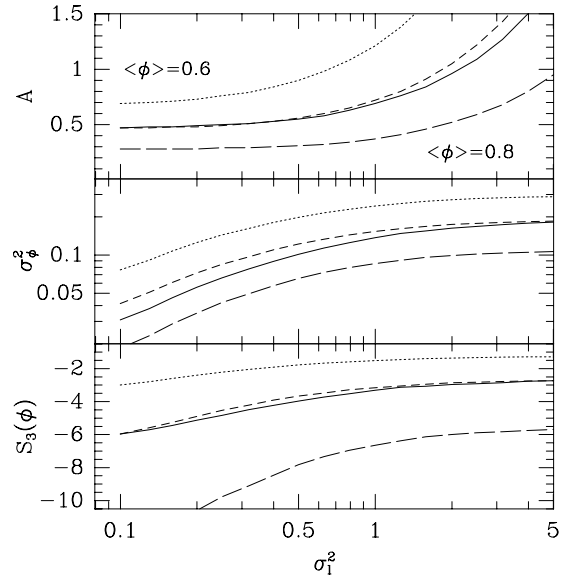}}
\caption[junk]{
Variation of the SC predictions (variance, middle panel and
skewness, bottom panel) with $\sigma_L^{2}$
for a fixed mean flux.
In the top panel we show the value of $A$ needed to give
mean flux $\langle\phi\rangle=0.6$ (dotted line), $0.7$ (short-dashed line), and
$0.8$ (long-dashed line).
 All results are in redshift-space (with thermal broadening) except for 
the continuous line which represents
results for 
$\langle\phi\rangle=0.7$ in real-space.
}
\label{fixm}
\end{figure}

We have seen in Fig. \ref{zflux} that  if we choose a specific value of $A$,
 the flux statistics can change fairly drastically if we add or remove the 
effects of peculiar velocities or thermal broadening.
When working with observational data, the value of $A$ is at best only known 
from estimates of the individual parameters in equation (\ref{afacs}),
$\Omega_{b}$, $\Gamma$, $H(z)$ to within a factor of 2. It would therefore
 be a good idea if we could fix $A$ directly  using \lya\ observations.
In principle, when working within the formalism we have adopted in this paper,
there are 4 unknown quantities, $\gamma$, $\sigma_{l}$, $A$ and $\beta$.
It has already been  found in Croft \etal (1998a) that a convenient
way of effectively determining the correct value of $A$ 
to use in numerical simulations is to choose the value which yields
the observed mean flux $\langle\phi\rangle$. We carry out the 
same procedure here,
so that when evaluating our predictions we make sure that 
 $\langle\phi\rangle$ 
is a fixed value. In Fig. \ref{fixm} we show results for three different 
values of $\langle\phi\rangle$, 0.6, 0.7, and 0.8, which are in the range 
measured from observations at $z=2-4$. In the top panel, we show the value of 
$A$  required for each value of $\langle\phi\rangle$, as a function of
$\sigma^{2}_{l}$. All results are in redshift-space, except for
$\langle\phi\rangle=0.7$, which we also show in real-space. We can see that once the mean 
flux is held fixed, the one-point moments of the flux  change only  little 
with the addition of redshift distortions.
In particular we find that the normalized hierarchical
 moments, $S_{3}(\phi)$ and
 $S_{4}(\phi)$ are practically identical in real and redshift-space.

\begin{figure} 
\centering
\centerline{\epsfysize=8.truecm 
\epsfbox{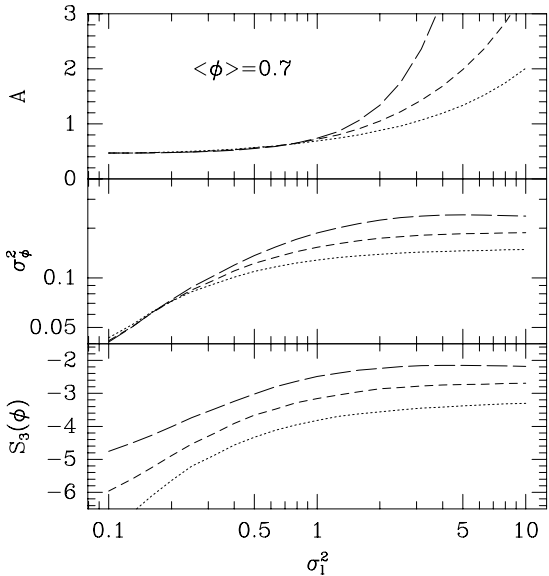}}
\caption[junk]{
Variation of the SC prediction
for the flux moments, with a fixed mean 
flux, as in Fig. \ref{fixm}, except this time for 
three different values of the
linear slope $\gamma$.
The dotted,  short-dashed and long-dashed
lines show the predictions for $\gamma=0,-1$ and $-2$
respectively. All results in 
redshift-space, and $\langle\phi\rangle=0.7$ for all curves.
}
\label{fix07g}
\end{figure}

In Fig. \ref{fix07g} we concentrate on $\langle\phi\rangle=0.7$ and show the
variation with $\gamma$, again in redshift-space. We can see that when 
the fluctuations in the underlying density field are large,
the variance and skewness tend to asymptotic values. As expected, more
negative values of $\gamma$, will produce spikier,
 more saturated absorption features and so give a larger variance and skewness.
To see the maximum values that this tendency
will produce, we can consider the fact that $\phi$
 can only lie between 0 and 1. There is therefore a limit to the level
 of clustering, given a specific value of the $\langle\phi\rangle$.
This will occur if a spectrum contains only pixels
which have either  $\phi=1$ or $\phi=0$ (either no absorption
or saturated). If the a fraction $f$ of the spectrum has $\phi=1$,
and the rest, a fraction $(1-f)$, has $\phi=0$, then $\langle\phi\rangle=f$.
Using the definition of $\sigma^{2}_\phi$ and
$S_{3}(\phi)$ in equations (\ref{eq:varf}) and (\ref{eq:s3f}),
we find that the variance in this case is 
\begin{equation}
\sigma^{2}_\phi=\frac{1}{f}-1=\frac{1}{\langle\phi\rangle}-1,
\end{equation}
and the normalized hierarchical skewness is given by
\begin{equation}
 S_{3}(\phi)=
\langle\phi\rangle\left(\frac{1}{\langle\phi\rangle}-1\right)
+\frac{(\langle\phi\rangle-1)}{([1/\langle\phi\rangle]-1)^{2}}.
\end{equation}
 For $\langle\phi\rangle=0.7$,
as in Fig. \ref{fix07g}, the maximum possible  $\sigma^{2}\phi=0.43$ 
and  $S_{3}(\phi)=-1.33$.

\bigskip

\section{Discussion}

The methods we have introduced in this paper are primarily
meant to be used as tools in the study of structure formation.
The predictive techniques for the
 one-point statistics should be most useful
when combined with information on clustering as a function of scale 
(two-point statistics), which we explore in a separate paper.
As far as applying our formalism to observations is concerned,
the one-point statistics we have discussed  should in principle
require data which resolves the structure in the forest, at least
at the level of the Jean's scale, or the thermal broadening width.
This would include Keck HIRES data (e.g., Kirkman \& Tytler 1997)
or other data with a spectral resolution better than $\sim10-15 \kms$.
Use of lower resolution data effectively involves smoothing along the
line-of-sight, and we leave the treatment of this anisotropic
smoothing window to future work.

At the simplest level, one could use our predictions for the one-point moments
to compare directly to observations. For example, given a measurement
of the mean flux (say $\langle\phi\rangle=0.7$), and a value for
 slope $\gamma$, one can use Fig. \ref{fix07g} to infer the 
predictions of gravitional instability for $S_{3}(\phi)$ and
$S_{4}(\phi)$, and then check them against the observed values.
This sort of test, while being suitable for checking 
whether predictions are generally compatible with Gaussian initial conditions,
and gravitational instability, is unlikely to be useful for discriminating
between popular Gaussian models, which have similar one point flux 
PDFs. For example, there is little difference
in the behaviour of the flux moments for
two models with different values of $\gamma$ shown in Figs. \ref{vssig1}
and \ref{vssig2}   (see also the small differences between $\Omega=1$ CDM and
$\Omega=0.4$ CDM in Rauch \etal 1997). 
We therefore 
advertise our one-point analytic predictions as being more suitable
for making wide searches of parameter space in order to ascertain the 
broad statistical trends expected in gravitational instability models
(for example, see below for the evolution of the moments with redshift).
Direct comparisons with observational data will be more fruitful
when carried out with two point statistics. We
will explore these, and how the analytic methods we
have developed here can be applied to them, in a future paper.
For the moment, several obvious applications of our 
techniques for making one-point predictions
suggest themselves, and we discuss these now. 
We also discuss the accuracy of the density evolution predictions, and compare
the present work to that of others.

\subsection{Non-Gaussian initial conditions}
There is a large parameter space of non-Gaussian PDFs to choose
from (see Fosalba, Gazta\~{n}aga \& Elizalde 1998). 
Here we will take a conservative approach and choose a model
with {\it mild} non-Gaussianities, with hierarchical correlations,
i.e., constant $S_J$, so that the cumulants $k_J \simeq k_2^{J-1}$
 They tend to the Gaussian result as  $k_2 \rightarrow 0$ more
quickly than
the dimensional scaling $k_J \simeq k_2^{J/2}$. As an illustrative example, we
will show results for a PDF based on the well known chi-squared 
distribution
(see e.g., Fosalba, Gazta\~{n}aga  \& Elizalde 1998 for details), also 
known as Pearson's Type III (PT3) PDF or Gamma PDF:
\beq
P(\rho) = {\rho^{1/\sigma^2-1} \over \Gamma(1/\sigma^2) 
(\sigma^2)^{1/\sigma^2}}
 \exp{\left(-{\rho\over{\sigma^2}}\right)},
\eeq
for $\rho\equiv 1+\delta \geq 0$. This PDF 
has $S_3=2$ and $S_4=6$ (in general $S_J=[J-1]!$). The number of the
chi-square degrees of freedom, $N$, in the  discrete version
of this distribution would
correspond to $N=2/\sigma^2$.
It is not difficult to build a non-Gaussian 
distribution with arbitrary values of $S_3$ or $S_4$, but to keep the
discussion simple we will just concentrate on the PT3 model. This model is not
only mildly non-Gaussian, in the sense of being hierarchical,
 but also has moderate 
values of $S_J$ (e.g., for comparison gravity produces $S_3=34/7$ from
 Gaussian initial conditions and unsmoothed fluctuations).

A possible motivation for introduction of the PT3 could be 
the isocurvature CDM cosmogony presented recently by Peebles (1998),
which has as initial conditions a one-point unsmoothed PDF given by
a chi-square distribution with $N=1$ (i.e. a PT3 with $\sigma^2=2$). Smoothing
would introduce higher levels of non-Gaussianities through the two-point
function, so this is a conservative approach.
A more generic motivation follows from the arguments presented
in Fosalba, Gaztanaga \& Elizalde (1998), who find that this distribution
plays a central role in non-Gaussian distributions that arise from
combinations of a Gaussian variables.

\begin{figure} 
\centering
\centerline{\epsfysize=8.truecm 
\epsfbox{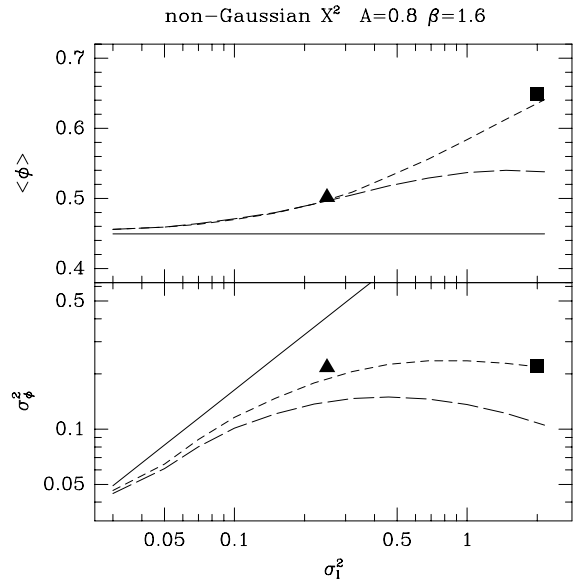}}
\caption[junk]{
A comparison of Gaussian and non-Gaussian models.
Mean flux $\langle\phi\rangle$ and
variance $\sigma^2_\phi$ (bottom) 
are plotted as a function of the linear
variance $\sigma^2_L$ for $\gamma=1$, $A=0.8$ 
and $\beta=1.6$. The short-dashed  and long-dashed
lines show the SC predictions for a Gaussian model
and a non-Gaussian PT3 model respectively.
Symbols corrspond to the Gaussian simulations, with the squares
representing set (a) (which has $\gamma\simeq1$)
and the triangles set (d) (which has $\gamma\simeq-1.5$) . 
 The continuous lines are the 
perturbative predictions  of Section 4.3
All results are in real-space.} 
\label{ngflux}
\end{figure} 

\begin{figure} 
\centering
\centerline{\epsfysize=8.truecm 
\epsfbox{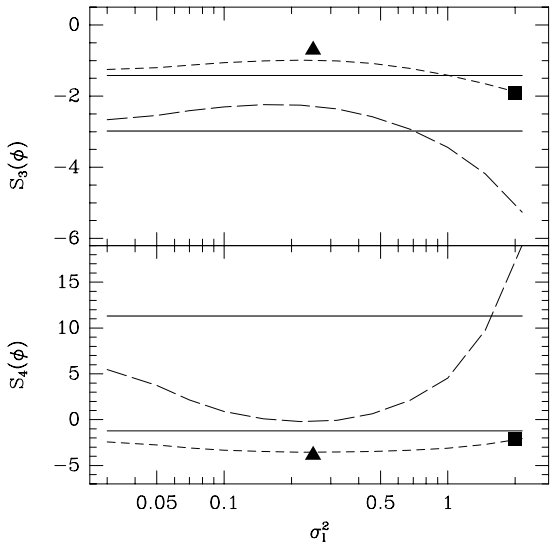}}
\caption[junk]{
A comparison of Gaussian and non-Gaussian models.
We plot the skewness $S_3$
and kurtosis $S_4$.
The symbols and line types are the same as in Fig. \ref{ngflux} 
 The continous lines are the 
perturbative predictions, which are different for the Gaussian
(thin line) and non-Gaussian (thick line) models.}
\label{ngs34}
\end{figure}

In Figures \ref{ngflux}-\ref{ngs34} we compare the Gaussian and 
non-Gaussian PT3
predictions. As can be seen, even in this mildly non-Gaussian model the 
predictions are quite different and can be clearly  distinguished from
the Gaussian simulations (symbols). 
The SC predictions for  Gaussian initial conditions
have been evaluated for $\gamma=-1$. 
The squares correspond to simulation set (a), which  
also have $\gamma\simeq-1$. The triangles are from set (c) and
have $\gamma=-1.5$. This value of $\gamma$ is different from
that used for the SC Gaussian predictions, although 
the simulation point for $\sigma^{2}_{\phi}$
 is still closer to them than the non-Gaussian predictions.

The leading order perturbative predictions, shown as 
continuous lines in the Figures, are the same for $\langle\phi\rangle$ and
$\sigma^2_\phi$, because the non-Gaussian model is hierarchical.
However for $S_3$ and $S_4$ the predictions are different.
It is easy to show that to leading order the (hierarchical) non-Gaussian
model predictions are given by  equation (\ref{eq:ptflux}) with:

\bea
S_3 &=& S_3(IC) + S_3(G) \\
S_4 &=& S_4(IC) + S4(G)+ 4 S_3(G) S_3(IC) \nn 
\label{eq:ptfluxng}
\eea
where  $S_3(IC)$ and
 $S_4(IC)$ are the initial conditions [$S_J(IC)=(J-1)!$ for PT3] and 
$S_3(G)$ and
 $S_4(G)$ are the leading order gravitational values: 
$S_3(G)=34/7+\gamma$ and $S_4(G)=60712/1323+62/3\gamma +7/3\gamma^2$.
As can be seen in Fig. \ref{ngs34} the perturbative values are only
reached asymptotically as $\sigma_L \rightarrow 0$.
The skewness $S_3$ is significantly lower in the PT3 model and
the kurtosis $S_4$ is much larger, changing from $S_4 \simeq -2$ in the 
Gaussian model 
to $S_4 \simeq 20$ in the non-Gaussian one.

\begin{figure} 
\centering
\centerline{\epsfysize=8.truecm 
\epsfbox{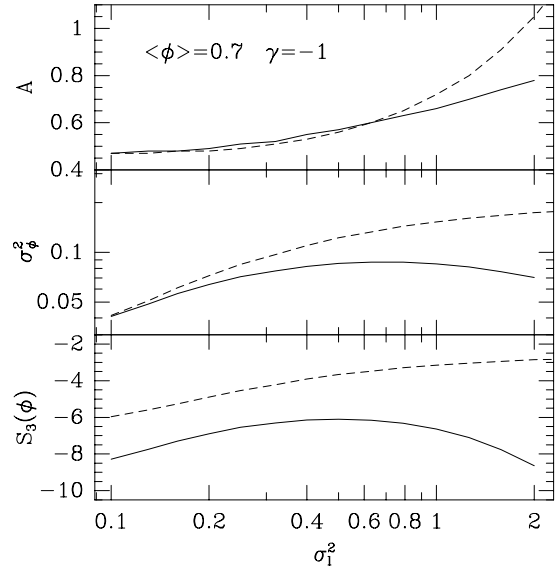}}
\caption[junk]{
Comparison between non-Gaussian and Gaussian predictions when $\langle\phi\rangle$ is
held fixed.
We plot the optical depth amplitude $A$ (top), variance $\sigma^2_\phi$
and skewness $S_3(\phi)$ (bottom) as a function of the linear
variance $\sigma^2_L$ for a mean flux $\langle\phi\rangle=0.7$, $\gamma=1$,  
and $\beta=1.6$. The short-dashed  and continuous 
lines show the predictions of the Gaussian and non-Gaussian PT3 models.
These results are in redshift-space.} 
\label{fixmflux07ng}
\end{figure} 

Fig. \ref{fixmflux07ng}, shows a comparion between 
the Gaussian and PT3 models, for  a fixed mean flux.
We noted previously (e.g., Fig \ref{fixm}) the useful fact that  
there is little difference between the 
predictions with or without redshift distortions 
when the mean flux is fixed in this way.
Fortunately, this is not the case for non-Gaussianities, as we can see here.
The results change considerably if we assume different initial conditions.

\subsection{Redshift evolution}

\begin{figure} 
\centering
\centerline{\epsfysize=8.truecm 
\epsfbox{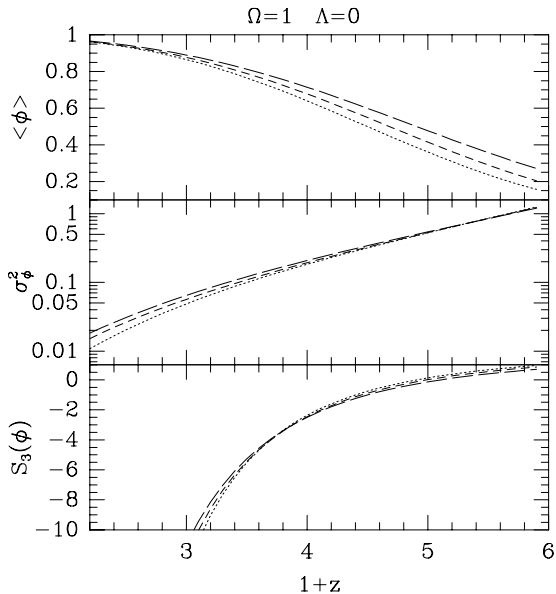}}
\caption[junk]{
The variation of the one-point statistics with redshift,
for three different values of $\gamma$, 0 (dotted line), -1 (short-dashed
line), and  -2 (long-dashed line).
The parameter $A=1.2$ at $z=3$ in all cases, and $\sigma^{2}_{L}=2$.
 Results are 
 for $\Omega=1$ and are 
in redshift-space with
thermal broadening.
}
\label{zevolg}
\end{figure}

We can see from equation (\ref{afacs}) 
that $A$ will change strongly with redshift,
and that there will be a dependence on 
cosmology through the variation in 
$H(z)$. We can investigate how the one-point flux statistics
 will vary using our formalism. We compute the evolution of $H(z)$ and 
$\sigma^{2}_{l}$ and show the results for $\Omega_{0}=1$ and three different
values of $\gamma$ in  Fig.
\ref{zevolg}. We can see that both $\sigma^2_{\phi}$ and
$S_{3}(\phi)$ become much smaller as the mean absorption falls.
In these figures we are assuming that $\gamma$ is not changing as a function
of redshift, or equivalently that the smoothing scale is
fixed in comoving $\hmpc$.

\begin{figure} 
\centering
\centerline{\epsfysize=8.truecm 
\epsfbox{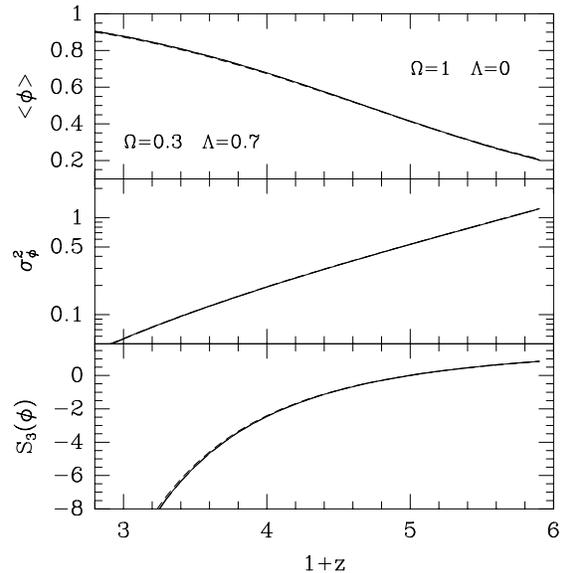}}
\caption[junk]{
The variation of the one-point statistics with redshift,
for three different cosmologies.
We have plotted results for $\Omega=1$ (solid line),
$\Omega_{0}=0.3, \Omega_{\Lambda}=0$ (short-dashed line),
and $\Omega_{0}=0.3, \Omega_{\Lambda}=0.7$ (long-dashed line).
In all cases, $\sigma^{2}_{L}=2$ and $\langle\phi\rangle=0.66$ at $z=3$.
Results are in redshift-space with thermal broadening.
}
\label{zevolc}
\end{figure}

\begin{figure} 
\centering
\centerline{\epsfysize=8.truecm 
\epsfbox{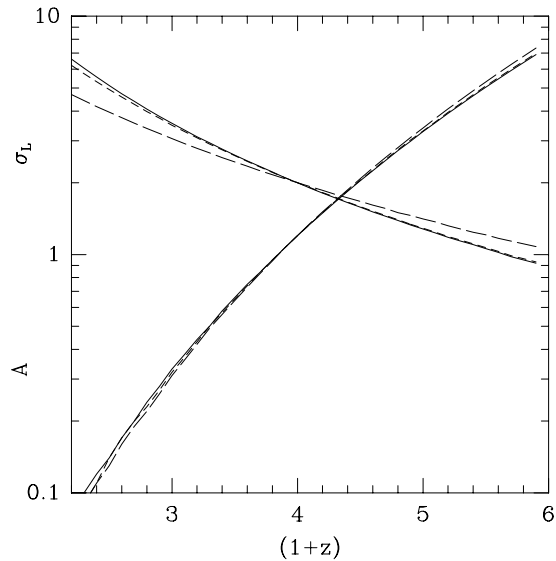}}
\caption[junk]{
Variation with redshift of $A$ and $\sigma^{2}_{l}$.
Results have been plotted for 3 differemt cosmologies,
$\Omega=1$ (solid line),
$\Omega_{0}=0.3, \Omega_{\Lambda}=0$ (short-dashed line),
and $\Omega_{0}=0.3, \Omega_{\Lambda}=0.7$ (long-dashed line).
}
\label{checkz}
\end{figure}

In Fig. \ref{zevolc} we plot the results for three different 
background cosmologies,  two flat models, one with $\Omega=1$
 and one with $\Omega_{0}=0.3$ and $\Omega_{\Lambda}=0.7$, as well
as an open model with $\Omega_{0}=0.3$. 
All models have the same $\gamma$, $\sigma_{l}$ and the same mean flux 
($\langle\phi\rangle=0.66$) at z=3. 
We can see that there is virtually
no visible dependence on cosmology in the plot.  The main
source of variation in $A$ which we
have not included, is the evolution of the photoionization 
rate $\Gamma$, which 
will change as the population of  sources for the
ionizing background changes. From 
Fig. \ref{zevolc} it is evident that inferences about the evolution
of the UV background should be fairly insensitive to the assumed cosmology.
We can see why this is so by looking at Fig. \ref{checkz}, where we
plot the evolution of $A$ and $\sigma^{2}_{l}$ separately.
If both quantities are fixed in the middle of the $z$ range as we have done,
then the changes over the range of validity of the FGPA ($z\simgt2$) are
small.

\subsection{The bias between flux and mass fluctuations}

\begin{figure} 
\centering
\centerline{\epsfysize=8.truecm 
\epsfbox{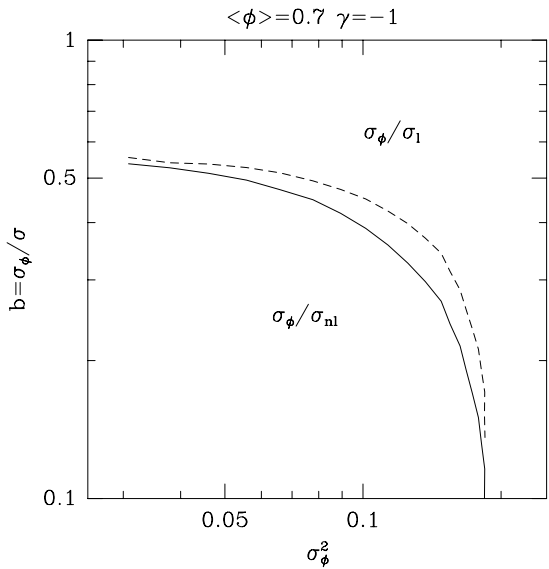}}
\caption[junk]{
The bias between flux and mass fluctuations. We show the bias
(see equation [\ref{bias}]) between the flux
and the linear mass ($\sigma_{l}$) as a dotted line and the bias
between the flux and the non-linear mass ($\sigma_{nl}$) as a solid line.
Both these quantities are plotted as a function of
the observable $\sigma_{\phi}^{2}$. The statistics of $\phi$ have
been computed in redshift space with thermal broadening,
for $\langle\phi\rangle=0.7$, $\sigma^{2}_{L}=2$ and $\gamma=-1$.
}
\label{biasc}
\end{figure}

In analogy with galaxy bias, we can define the bias of the flux
with respect to mass fluctuations as 
\begin{equation}
b=\sqrt{\frac{\sigma^{2}_{\phi}}{\sigma^{2}_{\rho}}}.
\label{bias}
\end{equation}
Unlike the case with galaxy bias, we can easily predict this quantity 
analytically using our formalism. We can choose between 
two sorts of bias, either the bias between the linear $\rho$, or the 
nonlinear  $\rho$. In Fig. \ref{biasc} we have plotted both of these
as a function of the variance in the flux, $\sigma_{\phi}$. As we are dealing
 with one-point statistics in this paper, we do not discuss the scale 
dependence of bias. However we will do so in Paper II in this series
 (Gazta\~{n}aga \& Croft 1999).
We should point out though that
in Croft \etal (1998a), it was found that
the shape of a two-point clustering statistic of the flux (in that case 
the power spectrum) follows well that of the linear mass.
The bias between the two was found in that paper by 
using a procedure which involved running numerical simulations set 
up with the power spectrum shape measured
from observations and comparing the clustering level in simulated
spectra with the observed clustering.
In this paper, we can find the bias level in a simpler fashion.
 We note that for
small values of the fluctuation amplitude, $b$ tends towards the values
predicted by perturbation theory (Section 4.3). For larger values, such
as those likely to be encountered in observations, a fully 
non-linear treatment,
such as the one presented here, is needed.

\subsection{Accuracy of the approximations for density evolution}

We have seen (e.g., Fig. \ref{pdf2}) that the predictions for the PDF 
have the right qualitative features
(as a function of $\gamma$ and $\sigma_L$) but do not reproduce it
 in all its details, even around  
$\delta=0$, because the SC is just a local approximation and does not
include shear.  The results of FG98 
indicate that the statistics of the weakly non-linear 
density moments are dominated by the local dynamical 
contribution to the evolution of the PDF, 
and shear forces are subdominant (they tend to 
cancel out when taking the mean). 
We find here  that a similar cancellation occurs when considering
the PDF of the flux, $\phi$,
even when $\sigma_L \simgt 1$.

Regarding the predictions for the 1-point moments of
the flux, we have checked that the 
Spherical Collapse (SC) model yields
almost identical results to the Generalized 
Zel'dovich Approximation (GZA), in Eq[\ref{GZA}],
with $\alpha=21/13$. This is true both in real and redshift-space,
and also holds for the prediction of the velocity divergence $\theta$.
This is an interesting result because
although the SC model is better motivated from the theoretical
point of view, the GZA model is much simpler to implement.
In particular the GZA model provides us with analytical
expressions for the PDF (i.e., equation [\ref{GZApdf}] and Fig. \ref{fpdf}), 
which can be used in practice to make the predictions.

As shown in Fosalba \& Gazta\~naga (1998b)
 the SC approach to modelling non-linearities 
does not work as well for $\theta$ as for $\rho$. In particular,
it was found that the next to leading order (or loop) 
non-linear corrections are not as accurate, indicating that 
tidal effects are more important for $\theta$.
 This could partially explain why the redshift distortion modelling
(see Fig. \ref{zflux}) requires the addition of an extra velocity
dispersion in order to match the results of simulations. 

\subsection{Comparison to other work}
The first attempts to constrain cosmology using the \lya\ forest 
focussed on comparisons between
simulated data generated with
 specific cosmological models
and observational data, using traditional line statistics. When 
the simulated and observed spectra are decomposed into a superposition of 
Voigt-profile lines,  the distribution of column densities of these lines and
the distribution of their widths (``b-parameters'') can be reasonably well
reproduced by gravitational instability models (e.g., Dav\'{e} \etal 1997,
although see Bryan \etal 1998).
 In the context of these models, the line parameters do
not have a direct physical meaning, as these statistics were intended to 
describe discrete thermally broadened lines. It is possible to 
use these traditional statistics to characterise the amount of small scale
power in the underlying density field, for example (see e.g.,
Hui, Gnedin \& Zhang, Gnedin 1998). However, statistics which are more
 attuned to the continous nature of the flux distribution and the underlying 
density field have advantages, as well as promising to be more sensitive
 discriminants, continous flux statistics can be designed to be less affected
 by noise and choice of technique than profile fitting.

As the modern view of the \lya\ is essentially an outgrowth
of structure formation theory, it makes sense
to borrow  statistical analysis techniques used in the study of the galaxy
 distribution. Unlike the galaxy distribution, however, the \lya\ forest
offers a truly continuous distribution of flux, with no shot noise (albeit in 
1 dimension), and a well motivated
theoretical relationship between the observed flux and the underlying mass.

So far, analysis of spectra using such
 continuous flux statistics has  mainly involved  
specific cosmological models, and direct comparison of simulations
with observations.
The mean flux, $\langle\phi\rangle$ is the most obvious
flux statistic to calculate. Its measurement from observations
has been carried out by several authors (e.g, Press, Rybicki \&
Scheider, 1993, Zuo \& Lu 1993), and there is an extensive
discussion in the literature about
 what is usually quoted as the mean flux decrement, 
$D_{A}=1-\langle\phi\rangle$, or the mean effective optical depth,
$\overline{\tau}_{\rm eff}=-ln\langle\phi\rangle$ 

 The probability 
distribution of the flux has been investigated by Miralda-Escude \etal (1996),
Croft \etal (1996), Cen (1997), Rauch \etal (1997), Zhang \etal (1998),
and Weinberg \etal (1998). Other statistics such as
the two point correlation function of the flux have been introduced
 (Zuo \& Bond 1994), 
the power spectrum of the flux (Croft \etal 1998a, Hui 1999,
the two point pdf of the flux (Miralda-Escud\'{e} \etal 1997),
 and the number of 
times a spectrum crosses a threshold per unit length (Miralda-Escud\'{e}
\etal 1996,
 Croft \etal 1996,  Weinberg \etal 1998).
 Methods have been developed to reconstruct 
properties of the underlying mass distribution, such as the matter power
spectrum (Croft \etal 1998a, Hui 1999,
 using our theoretical assumptions
for the relation between mass and flux.  A technique for carrying out a direct 
inversion from the flux to the mass distribution has been described by Nusser 
and Haehnelt (1998).
In the present paper, we emphasise the use of statistics which have
been used extensively in the study of galaxy clustering, in particular the 
higher order moments (e.g., Gazta\~{n}aga \& Frieman 1994).
 These statistics, and their behaviour
when used to quantify the evolution of density perturbations in the
quasil-linear regime have been the subject of much attention. It would seem 
that extending their use to the study of the \lya\ forest may offer us a 
good chance to combine our knowledge of gravitional instability with that of
the IGM and in doing so make progress in both disciplines. 

On the predictions side, many pieces of
 analytic work have been carried out which incorporate the dominant 
physical processes involved in producing high-redshift \lya\ absorption
(processes summarized in Section 2).
 The studies' 
most important differences have been
 in the schemes used to follow the evolution
of density perturbations. These have included linear theory (Bi 1993,
Bi, Ge \& Fang 1995)
 the lognormal approximation (Gnedin \& Hui 1996, Bi \& Davidsen 1997),
and the Zel'dovich Approximation (McGill 1990, 
Reisennegger \& Miralda-Escud\'{e} 1995, Hui, Gnedin \& Zhang 1997).
Unlike these approximations, the SC model used in this paper
 is able to reproduce exactly the perturbation theory results for
clustering. This accuracy makes it useful for calculating
high-order statistics of the flux, in our search for the signatures
of gravitational instability. It is important to realize that
we have not used the SC model to make simulated QSO \lya\ spectra,
but that we have used its predictions for the properties
of the mass to predict the
statistics of the \lya\ flux. With such an approach
(similar to that taken by Reisennegger \& Miralda-Escud\'{e} 1995)
 we can quickly and easily vary parameters in order
to explore for example the dependence of a particular statistic on redshift
(Section 6.2). Some tasks which previously required numerical simulations,
such as finding the bias between density and
flux fluctuations (e.g., Croft \etal 1998ab), can be
carried out analytically.  

The fully non-linear analysis we have described in this paper will allow
 one to carry out many analyses of clustering where the precise
relationship  between the statistics of the mass and the flux is
important. This includes attempts to constain the cosmic geometry
from the clustering measured between adjacent QSO lines of sight (e.g.,
McDonald \& Miralda-Escud\'{e} 1999 Hui, Stebbins \& Burles 1999.
In such situations, a non-linear theory of redshift distortions in the
\lya\ forest and of the bias between the fluctuations in the
observed flux and the mass, such as we have presented in this paper 
should be very useful.

\section{Summary and conclusions}

We have presented a fully non-linear analytical treatment of the
one-point clustering properties of the high-z \lya\ forest in the gravitational
instability scenario.
The formalism we have presented should prove to be a
useful tool for studying the forest, and has immediate application to 
the calculation of two-point statistics (see Paper II).
The two main ingredients we have used  
are the Spherical Collapse model (SC) or
shear-free approximation for the evolution of density
perturbations, and the Fluctuating Gunn-Peterson Approximation
for the relation between density and \lya\ optical depth.
The predictions for the one-point clustering of the mass made using the
SC model depend only on two 
parameters, $\sigma^{2}_{L}$ and $\gamma$. 
These are, respectively, the linear variance of density
fluctuations, and the local slope of the linear correlation function.
 In the FGPA, the relation between the mass distribution
and \lya\ forest optical depth 
is largely governed by one parameter, $A$, which
can be set by appealing to observational
measurements of the mean flux.
 The predictions of the SC model for the density are 
typically quite non-linear ($\sigma^{2} \sim 2$ or more).
While these predictions are not expected to be accurate for the 
high density tail of the distribution,
the weighting of the FGPA relation means that the 
statistics of the flux are governed by the (quasi-linear) 
density regime where the SC ${\it is}$ accurate.
The \lya\ forest is therefore well suited to study using 
such an approximate analytical technique.

We note that the analytical predictions can be used in tests
of the picture of \lya\ forest formation and
the applicability of the FGPA.
With the extra information afforded by the two-point statistics and
considering the evolution of clustering as a function of redshift,
it will be possible to look for the signatures of any deviation from the
theroretical picture. Consistency tests for the gravitational instability
scenario include checking the evolution of the moments as 
a function of redshift, the scaling of the hierarchical moments,
and their dependence on scale.

We plan to use our predictive techniques in future work
to extract information from observations, using both one-point and two-point
statistics. In the present paper, we have concentrated on developing 
an analytical framework for studying the clustering of the 
transmitted flux 
in the  forest region of QSO spectra.
 Some of the results of our present exploration 
of one-point statistics include the following:

\noindent $\bullet$ Using our formalism we are able to 
 estimate the bias between mass and flux fluctuations
without resorting to simulations.\\
$\bullet$ We can make predictions for the clustering properties of the \lya\ 
forest flux in both Gaussian and non-Gaussian models. We find large
 differences between the two in an example case.\\
$\bullet$ In the limit of small fluctuations, our non-linear analytical
treatment converges to the same
results as those from Perturbation Theory calculations.\\ 
$\bullet$ For larger fluctuations, where Perturbation Theory is
no longer valid, we find our treatment to give accurate results 
compared to statistics evaluated from 
 N-body simulated spectra. These predictions are most accurate
when the linear variance of the density field is $\sim 4$
and below. For values above this, the qualitative
behaviour of the high order moments is reproduced.\\
$\bullet$ We can follow the evolution of the one-point statistics
of the flux as
a function of redshift. We find that the difference between predictions 
for different cosmologies is small, so that comparison with
observations should be useful in constraining the evolution
of the ionizing background intensity.\\
$\bullet$ If we normalise our predictions so that the mean flux
is held fixed (for example to the observed value), we find that the 
statistics of the flux are relatively insensitive to the effects of
redshift distortions induced by peculiar velocities or 
thermal broadening. This is most valid for the higher order normalised
hierarchical moments.\\

The high-$z$ \lya\ forest is amenable to study using analytical treatments,
a fact which gives it great value as a probe of structure formation.
Application of analytical theory to measurements
from the  many observational datasets currently available promises
to reveal much, both about the validity of our theoretical assumptions, and 
about cosmology.

\vspace{1cm}

{\bf Acknowledgments}

We thank George Efstathiou for the use of his P$^{3}$M N-Body code.
We also thank Jasjeet Bagla and Pablo Fosalba
 for useful discussions, and the anonymous referee for suggestions
which improved the paper.
EG acknowledges support from 
CSIC, DGICYT (Spain), project
PB93-0035, and CIRIT, grant GR94-8001 and
1996BEAI300192. 
RACC acknowledges support from NASA Astrophysical Theory Grant NAG5-3820
and CESCA for support during a visit to the IEEC.

\section{References}
\def\refe {\par \hangindent=.7cm \hangafter=1 \noindent}
\def\apj { ApJ }
\def\aap {A \& A }
\def\ajs{ ApJS }
\def\aj{AJ}
\def\apjs{ ApJS }
\def\mnras { MNRAS }
\def\apjl { Ap. J. Let. }

\refe Baugh, C.M., Gazta\~{n}aga, E., \& Efstathiou, G.,
1995, \mnras, 274, 1049

\refe Bi, H.G., Boerner, G., Chu, Y., 1992, \aap, 266, 1

\refe Bi, H.G. 1993, \apj, 405, 479

\refe Bi, H., Ge, J., \& Fang, L.-Z. 1995, \apj, 452, 90

\refe Bi, H.G. \& Davidsen, A. 1997, \apj , 479, 523 

\refe Bechtold, J., Crotts, A. P. S., Duncan, R. C., \& Fang, Y. 1994,
 \apj, 437, L83

\refe Bernardeau, F., 1992, ApJ, 392, 1 

\refe Bernardeau, F., 1994, A\&A, 291, 697

\refe Bernardeau, F., \& Kofman, L., 1995, \apj, 443, 479 

\refe Bryan, G.L., Machacek, M., Anninos, P., Norman, M.L., 1998,
\apj, 517, 13

\refe Cen, R., 1997, 479, L85

\refe Cen, R., Miralda-Escud\'e, J., Ostriker, J.P., \& Rauch, M. 1994,
 \apj, 437, L9

\refe Croft, R.A.C., \& Efstathiou, G., 1994, \mnras, 267, 390  

\refe Croft, R.A.C., Weinberg, D.H., Hernquist, L. \& Katz, N. 1996,
In : ``Proceedings of the 18th Texas Symposium on Relativistic 
Astrophysics and Cosmology'', eds. Olinto, A., Frieman, J., 
\& Schramm, D.N.

\refe Croft, R.A.C., Weinberg, D.H., Katz, N., \& Hernquist, L. 1997,
\apj, 488, 532

\refe Croft, R.A.C., Weinberg, D.H., Katz, N., \& Hernquist, L. 1998a,
\apj, 495, 44

\refe Croft, R.A.C., Weinberg, D.H., Pettini, M., 
Hernquist, L., \& Katz, N. 1999, \apj 520, 1

\refe Crotts, A.P.S., \& Fang, Y., 1998, \apj, 497, 67

\refe Dav\'e, R., Hernquist, L., Weinberg, D. H., \& Katz, N. 1997,
 \apj, 477, 21

\refe Dav\'e, R., Hernquist, L., Katz, N., \& Weinberg, D. H.,  1999,
 \apj, 511, 521

\refe Dinshaw, N., Impey, C. D., Foltz, C. B., Weymann, R. J., \&
Chaffee, F. H. 1994, \apj, 437, L87

\refe Dinshaw, N., Foltz, C. B., Impey, C. D., Weymann, R. J., \&
Morris, S. L. 1995, Nature, 373, 223

\refe Efstathiou, G.,  \& Eastwood, J.W., 1981, \mnras, 194, 503

\refe Efstathiou, G., Bond, J.R., \& White, S.D.M., 1992, \mnras, 
258, 1p

\refe Efstathiou, G., Davis, M., White, S.D.M., \& Frenk, C.S.,
1985, \apjs, 57, 241

\refe Fardal, M., \& Shull, M., 1993, \apj, 415, 524

\refe Fosalba, P., \& Gazta\~naga, E., 1998a, \mnras, 301, 503
(FG98)
\refe Fosalba, P., \& Gazta\~naga, E., 1998b, \mnras, 301, 535

\refe Fry, J.N., \& Gazta\~naga, E., 1993, \apj, 413, 447

\refe Gazta\~naga, E., \& Baugh, C.M., 1995, MNRAS, 273, L5

\refe Gazta\~{n}aga, E., \& Croft, R.A.C., 1999, {\it in preparation}
(Paper II)

\refe Gazta\~naga, E., \& Fosalba, P., 1998, \mnras, 301, 524
astro-ph/9712095

\refe Gazta\~naga, E., \& Frieman, J.A., 1994, MNRAS, 425, 392

\refe Gnedin, N. Y. \& Hui, L. 1996, \apj, 472, L73

\refe Gnedin, N. Y., \& Hui, L. 1998, \mnras, 296, 44

\refe Gunn, J.E., \& Peterson, B.A. 1965, \apj, 142, 1633

\refe Haiman, Z., Thoul, A., \& Loeb, A., 1996, \apj, 464, 523

\refe Hernquist L., Katz N., Weinberg D.H., \& Miralda-Escud\'e J. 
1996, \apj, 457, L5

\refe Hockney,R.W., \& Eastwood, J.W., 1988, Computer Simulation
Using Particles, Bristol: Hilger

\refe Hu, E.M., Kim, T-S., Cowie, L.L., Songaila, A.,\& Rauch, M.,
1995, \aj, 110, 1526

\refe Hui, L., 1999 \apj, 516, 519

\refe Hui, L., \& Gnedin, N. 1997, \mnras, 292, 27

\refe Hui, L., Gnedin, N., \& Zhang, Y. 1997, \apj, 486, 599

\refe Hui, L., Stebbins, A., \& Burles, S., 1999, \apj, 
611, L5

\refe Juszkiewicz, R., Weinberg, D.H., Amsterdamski, P., Chodorowski, M.,
\& Bouchet, F., 1995, ApJ, 442, 39

\refe Kaiser, N., 1987, MNRAS, 227, 1

\refe Katz, N., Weinberg D.H., \& Hernquist, L. 1996, \apjs, 105, 19

\refe Kim, T-S, Hu, E.M., Cowie, L.L., Songaila, A., 1997, \aj, 114,1

\refe Kirkman, D., \& Tytler, D., \apj, 484, 672

\refe Kofman, L., Bertschinger, E., Gelb, J.M., Nusser, A., \& Dekel, A.,
1994, ApJ, 420, 44

\refe Lu, L., Sargent, W.L.W., Womble, D., \& Takada-Hidai, M.,
1996, \apj, 472, 509

\refe Lynds, C. R., 1971, ApJ, 164, L73

\refe McDonald, P. \& Miralda-Escud\'e, J. 1998, \apj, 518, 24

\refe McGill, C. 1990, \mnras, 242, 544

\refe Miralda-Escud\'e J., Cen R., Ostriker J.P., \& Rauch M. 1996,
 \apj, 471, 582

\refe Miralda-Escud\'e J.,  Rauch, M., Sargent, W., Barlow, T.A.,
Weinberg, D.H., Hernquist, L., Katz, N., Cen, R., \& Ostriker, J.P., 1997,
in: ``Proceedings of 13th
     IAP Colloquium: Structure and Evolution of the IGM from QSO 
Absorption Line Systems'', eds. Petitjean,P., \&   Charlot, S., 

\refe Miralda-Escud\'e J., \& Rees, M. J. 1994, \mnras, 266, 343

\refe Munshi, D., Sahni, V., \& Starobinsky, A.A., 1994, ApJ, 436, 517

\refe Nusser, A., \& Haehnelt, M., 1999, \mnras, 300, 1027

\refe Peebles, P.J.E., 1980, The Large-Scale Structure of
the Universe.  Princeton Univ. Press, Princeton

\refe Peebles, P.J.E., 1993, Principles of Physical Cosmology,
Princeton Univ, Press, Princeton

\refe Peebles, P.J.E., 1999, \apj, 510, 531

\refe Press W.H., Rybicki G.B., Schneider D.P., 1993, \apj, 414, 64 

\refe Protogeros, Z.A.M., \& Scherrer, R.J. 1997, MNRAS, 284, 425 

\refe Rauch, M., Miralda-Escud\'e, J., Sargent, W. L. W., Barlow, T. A.,
Weinberg, D. H., Hernquist, L., Katz, N., Cen, R., \& Ostriker, J. P.
1997, \apj, 489, 7

\refe Rauch, M., 1998, ARAA, in press

\refe Reisenegger, A. \& Miralda-Escud\'{e}, J. 1995, \apj, 449, 476

\refe Sargent, W.L.W., Young, P.J.,  Boksenberg, A., Tytler, D. 1980,
\apjs, 42, 41

\refe Scherrer, R., \& Gazta\~naga, E., 1998, in preparation

\refe Theuns, T., Leonard, A., Efstathiou, G., Pearce, F.R., Thomas, P.A.,
1998, \mnras, 301, 478

\refe Wadsley, J. W. \& Bond, J.R. 1996,
in  ``Computational Astrophysics'', 
Proceedings of the 12th Kingston Conference,
eds. Clarke, D., West, M., PASP, astro-ph 9612148

\refe Weinberg, D.H., Hernquist, L., Katz, N., Croft, R., \&
Miralda-Escud\'{e}, J., 1998a, In : Proc. of the 13th IAP Colloquium, 
Structure and Evolution of the IGM from QSO
Absorption Line Systems, eds. Petitjean, P., \&  Charlot, S.,
Nouvelles Fronti\`eres, Paris, astro-ph/9709303

\refe Weinberg \etal, 1998b, In: 
Proceedings of the MPA/ESO Conference
"Evolution of Large Scale Structure: From Recombination to Garching",
astro-ph/9810142

\refe Zhang Y., Anninos P., \& Norman M.L. 1995, \apj, 453, L57

\refe Zhang Y., Meiksin, A.,  Anninos P., \& Norman M.L. 1998, \apj, 495, 63

\refe Zuo, L. 1992, \mnras, 258, 36

\refe Zuo, L., \& Bond, J.R., 1994, \apj, 423, 73

\refe Zuo, L., \& Lu, L., 1993, \apj, 418, 601

\appendix

\section{Non-linear mapping relations}

\subsection{Unsmoothed relations}

\subsubsection{The density}

For small linear fluctuations, a generic mapping 
can be  expressed in a Taylor series:

\beq
\rho(x) = \calG[\delta_L(x)] = \sum_k {\nu_k\over{k!}} 
\delta^k_L(x)
\label{taylor}
\eeq
where $\nu_k$ are the coefficients that define the transformation
in this limit. The above mapping can be used to predict the PDF
of the evolved field, or directly predict its cumulants. For example,
for the skewness we have:
\beq
\langle\de^3\rangle_c = \langle(\calG-\langle\calG\rangle)^3\rangle
= 3 \nu_2 \sigma_L^4 + \Or(\sigma_L^6)
\eeq
where we have made use of the Gaussianity of
$\delta_L$. In general we have:
\beq
\langle\de^J\rangle_c ~=~ S_J~ \sigma_L^{2(J-1)} + \Or(\sigma_L^{2J}).
\eeq
For the skewness $S_3=3 \nu_2$, and in general
$S_J$ can be given in terms of $\nu_k$ 
(e.g., see FG98 or the original work by Bernardeau 1992, which is in
terms of the generating functionals). 

For the SC (see below) we have:
\bea
\nu_2 &=& {34\over 21} \sim 1.62 \nn \\ 
\nu_3 &=& {682\over 189} \sim 3.61 \nn \\ 
\nu_4 &=& {446440\over 43659} \sim 10.22 \nn \\
\nu_5 &=& {8546480\over 243243} \sim 35.13
\label{nusc}
\eea
These numbers give the correct leading PT contribution to
the Jth-order cumulants $\langle\de^J\rangle_C$ for Gaussian IC,
e.g., $S_3=34/7$ (Peebles 1980, \S42).
The important point to notice
here is that although the local mapping is not the exact
solution to the evolution of $\delta$ (which is in general
non-local), it gives the correct clustering
properties in the weakly non-linear regime.
This is because of the symmetry involved in taking
the ensemble average $\langle...\rangle$ (Bernardeau 1992). 
This is also a good approximation for next to leading
terms (FG98) and also for non-Gaussian initial
conditions (Gazta\~naga \& Fosalba 1998). 

\subsubsection{Velocity divergence}

The density mapping can be used to predict the velocity 
divergence, defined here as:
\beq
\theta= {1\over{H}} \, \nabla \cdot v 
\eeq
where $v$ is the peculiar velocity field.
We can use the continuity equation:
\beq
{d\rho\over{dt}} +\,H\,\rho \, \theta =0
\eeq
to express $\theta$ as a Lagrangian mapping of $\delta_L$:
\beq
\theta=\calG_v[\delta_L]= - {1\over{H\,\rho}} \, {d\rho\over{dt}}
= -f_\Omega \,\, {\delta_L\over{\rho}} {d\rho\over{d\delta_L}},
\label{theta}
\eeq
where $f_\Omega \equiv d \ln D/d \ln a$, comes from applying the
chain rule to the derivative
of the linear growth factor: $\delta_L = D(t) \delta_{IC}$. 

\subsubsection{The SC model}

\begin{figure} 
\centering
\centerline{\epsfysize=8.truecm 
\epsfbox{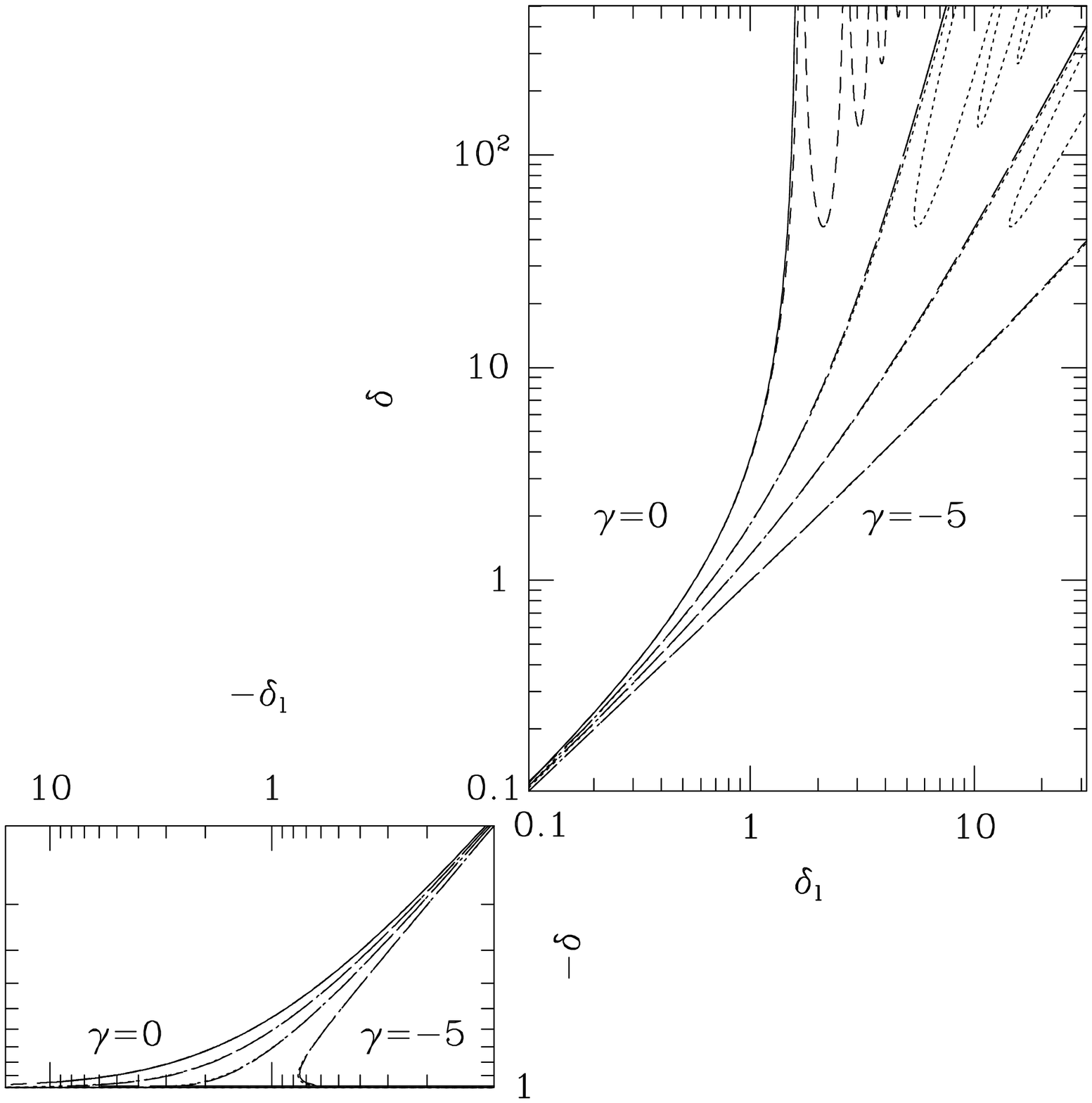}}
\caption[junk]{Local mappings for positive (top right-hand corner) and
negative fluctuations (bottom left-hand corner). The short-dashed line 
shows the (unsmoothed $\gamma=0$) 
spherical collapse (SC) result, which has a first bounce at
$\delta_l \simeq 1.69$. The
continuous line (which almost covers the short-dashed line)
is the GZA with $\alpha=21/13$ dimensions, which
becomes singular at $\delta_l \simeq 1.62$ and has its 
first two derivatives equal to those of SC.
 The three long dashed lines below
show the same GZA smoothed with increasing smoothing
indices of $\gamma=-1.5, -3, -5$. These curves are difficult
to distinguish from the corresponding SC mappings smoothed
with the same $\gamma$, which are displayed as dotted-lines.}
\label{scmapping}
\end{figure}

For large $\rho$ the spherical
collapse mapping $\delta=\calG[\de_L]$
can only be expressed in a parametric form through an
auxiliary variable $\psi$:

\begin{eqnarray}
\label{SC1}
\rho^{+} &=& {9 \over 2}{(\psi-\sin\psi)^2 \over
(1-\cos\psi)^3}, \nonumber\\
\delta_L^{+} &=& {3\over 5} [{3\over 4}(\psi-\sin\psi)]^{2/3},
\end{eqnarray}
for $\delta_L > 0$, which we call $\delta_L^{+}$, and

\begin{eqnarray}
\label{SC2}
\rho^{-} &=& {9 \over 2}{(\sinh\psi-\psi)^2 \over
(\cosh\psi-1)^3}, \nonumber\\
\delta_L^{-} &=& - {3\over 5} [{3\over 4}(\sinh\psi-\psi)]^{2/3},
\end{eqnarray}

for $\delta_L < 0$, which we call $\delta_L^{-}$. 
This parametric form complicates the 
analytical predictions,
but it can be  implemented
numerically. For $\psi=2 \pi$ the transformation becomes
singular and $\rho$ and $\de$ diverge. This is the first collapse
which occurs at $\delta_L=3/5(3\pi/2)^{2/3} \simeq 1.686$, 
as illustrated in Fig. \ref{scmapping}. It then bounces back
and recollapses in a periodic fashion. 

\subsubsection{The GZA model}

Another model we consider for the local transformation is the
{\it Generalized Zel'dovich Approximation} (GZA):
\beq
\rho= 1+\de = \left| 1 -{\delta_L\over{\alpha}}\right|^{-\alpha},
\label{GZA}
\eeq
which is a symmetric version of the ZA in $\alpha$ dimensions and
was also considered by Protogeros and Scherrer (1997). 
The case $\alpha=3/2$ was introduced by Bernardeau (1994) 
as a good approximation
to the SC model in the weakly non-linear regime. This 
value of $\alpha$ gives
$\nu_2 \simeq 1.67$ in equation (\ref{taylor}), compared to
the exact value $\nu_2=34/21 \simeq 1.62$.
Here we will concentrate instead on the case 
$\alpha=21/13$, as it reproduces
exactly the SC result to second order: $\nu_2=34/21$,
and therefore gives the exact skewness, $S_3=34/7$. 
This value of $\alpha$ also is a better
 approximation  than $\alpha=3/2$ around 
the singularity in $\de$, which occurs for $\de_L=\alpha \simeq 1.62$,
closer to $\delta_L \simeq 1.689$ than $\de_L=\alpha=1.5$. 
The higher order derivatives
for the case $\alpha=21/13$
are $\nu_3 \simeq3.62$, $\nu_4 \simeq 10.35$ and
$\nu_5= 35.99$, which should be compared to the
SC values in equation (\ref{nusc}). Another model of
interest is $\alpha=3/5(3\pi/2)^{2/3} \simeq 1.686$
which reproduces the behaviour exactly around the singularity, 
and gives $\nu_2 \simeq 1.59$, which is closer to the PT
value in the SC  than $\alpha=3/2$.

The velocity divergence, equation (\ref{theta}),  in the GZA 
has the simple form:
\beq
\theta = -f_\Omega  \,\, \de_L \, \rho^{1/\alpha}
= -f_\Omega \,\, \de_L \, \left| 1 -{\delta_L\over{\alpha}}\right|^{-1}
\label{thetagza}
\eeq

An important problem with these relations is that there is more than
one value of $\delta_L$ for a given value of $\delta$. 
This might be
important when it comes  to obtaining the PDF of $\delta$.
In practice this branching problem can be solved by assigning
to $\delta$ the probability associated with the different branches in
$\delta_L$. 
However, we have checked that this is not important 
for smoothed fields and this allows us to ignore this branching problem from
now on.

\subsection{Smoothing effects}

The above local relations correspond to unsmoothed
fluctuations.  We will focus instead
on fields smoothed with a top-hat filter, defined as, 
\bea
W_{TH} (\x,R) &=& 1 \quad if \quad |\x| \leq R , \nn  
\label{svar}
\eea
and zero otherwise, where $R$ is the smoothing radius.
 The volume corresponding 
to radius $R$ is  $V=4\pi/3 R^3$.
For  top hat smoothing, the  statistical properties of  the
smoothed fields can be obtained using the following prescription
(see also Bernardeau 1994, FG98). Consider an initial (Lagrangian) fluctuation
of infinitesimal mass $m_0$ extending over some volume 
$V_0= 4\pi/3 R_0^3$.
The statistics of the IC must be given as an input to estimate the corresponding 
evolved, non-linear, values after gravitational growth. This is also true for the
smoothed case, so that the shape and amplitude of the initial variance 
must be given or set 'by hand'. 
Consider first the case were we want the initial fluctuations
smoothed within a radius $R$, to have a power-law variance: 
\beq
\hat\sigma^2 = \sigma_0^2 \left({R\over{R_0}}\right)^\gamma,
\eeq
where $R_0$ and $\sigma_0^2$ relates to the ``unsmoothed'' amplitude. 
In the linear regime, this sets the amplitude of linear fluctuations: 
\beq
\hat\delta_L = D(t) \, 
\delta_0 \, \left({R\over{R_0}}\right)^{\gamma/2} 
 = \, \delta_L \, \left({R\over{R_0}}\right)^{\gamma/2}, 
\eeq
where $D(t)$ is the gravitational growth factor and $\delta_0$ is some
(unsmoothed) initial seed whose variance is  $\sigma_0^2$, so that
in the limit $R \rightarrow R_0$ we have 
$\hat\sigma^2_L \rightarrow D^2 \sigma_0^2 = \sigma^2_L$.

On the other hand, in the local picture of evolution, 
a given fluid element $m_0$ is isolated, so that for 
a mean density $\overline{n}$, the smoothed overdensity is:
\beq
\hat\rho = {m_0\over{\overline{n} V}} =  \left({R_0\over{R}}\right)^3,
\label{srho}
\eeq
as the IC are perfectly homogeneous ($m_0=\overline{n} V_0$), and the
mean density $\overline{n}$ does not change with time.

Puting the above equations together we have: 
\beq
\hat\rho = \left({\delta_L\over{\hat\delta_L}} \right)^{6/\gamma},
\label{srhodl}
\eeq 
which shows that smoothing acts like a (implicit) Lagrangian mapping.
Thus, given the unsmoothed mapping $\rho=\calG(\delta_L)$,
we can find the corresponding smoothed mapping by
solving the implicit relation:

\beq
\hat{\rho} =\calG\left[\,\hat\rho^{\gamma/6}\hat\de_L\,\right],
\label{smap}
\eeq
where $\gamma$ is the slope of the initial or linear
smoothed variance:
\beq
\gamma= {d\log{\sigma^2_L}\over{d\log{R}}}.
\label{slopedef}
\eeq
This can be easily generalized to non power-law relations
and reproduces the original result of Bernardeau (1994).

The velocity divergence can be thought
as a different $\delta_L$ mapping, i.e., equation (\ref{theta}). Thus
the smoothed results can be obtained as with equation (\ref{smap}):

\beq
\hat\theta=\calG_v[\hat\delta_L\hat\rho^{\gamma/6}].
\label{thetas}
\eeq

\subsubsection{Smoothing in the SC model}

In principle, the smoothing in $\rho$ is difficult to implement 
analytically for the SC, because $\calG$ is given by the implicit relations 
Eqs.[\ref{SC1}]-[\ref{SC2}]. However, the smoothing can be 
carried out easily in terms 
of the smoothed $\delta_L$, simply by noticing that 
from the above relation (equation [\ref{srhodl}]) we have:
\beq
\hat\delta_L[\psi] = {\delta_L[\psi]\over{\rho[\psi]^{\gamma/6}}},
\label{hatdel}
\eeq
where $\psi$ is such that $\hat\rho = \rho[\psi]$.
The smoothed PDF of $\hat\rho$ will be given in terms of
the smoothed PDF of the IC, which for a Gaussian field is
also a Gaussian. So we just have, following  Eq[\ref{nlpdf}]:
\beq
P(\hat\rho) \propto
{P_{IC}(\hat\de_L)\over{\hat\rho}} \left|{d\hat\de_L\over{d\hat\rho}}\right|
= 
{P_{IC}(\hat\de_L[\psi])\over{\rho[\psi]}} 
\left|{d\hat\de_L\over{d\psi}}\right|
\left|{d\rho\over{d\psi}}\right|^{-1},
\eeq
with $\hat\de_L[\psi]$ given by equation (\ref{hatdel}).
Care must be taken to use the correct relation, equation (\ref{SC1})
or equation (\ref{SC2}), depending on the sign of
$\delta_L$. For example, to estimate the moments of
$\rho$ we have:
\bea
\langle\rho^J\rangle &\equiv& \int P(\hat\rho) \, \hat\rho^J \, d\hat\rho \nonumber \\ 
&=& {1\over{N}}\int {P_{IC}(\hat\de_L^{+}[\psi])\over{\rho^{+}[\psi]}} 
\left|{d\hat\de_L^{+}\over{d\psi}}\right| \, \rho^{+}[\psi]^J \, d\psi
\, + \\ 
&+&
{1\over{N}}\int {P_{IC}(\hat\de_L^{-}[\psi])\over{\rho^{-}[\psi]}} 
\left|{d\hat\de_L^{-}\over{d\psi}}\right| \, \rho^{-}[\psi]^J \, d\psi
\, \nonumber,
\eea
where $N$ is the normalization factor, so that $\langle1\rangle=1$.
The superscript $^+$ denotes our use of equation (\ref{SC1}) 
and $^-$ our use of  equation (\ref{SC2}).

\subsubsection{Smoothing in the GZA model}

From now on, unless stated otherwise,
we use  $\de$ and  $\rho$
to refer to the smoothed fields. The unsmoothed case corresponds
to $\gamma=0$.
The smoothed GZA mapping is given implicitly by equation (\ref{smap}),
which for the GZA case of equation (\ref{GZA}) yields:
\beq
\rho= \left|1- {\de_L \, \rho^{\gamma/6}\over{\alpha}} \right|^{-\alpha} 
\eeq
which nicely maps $\de_L \in [-\infty,\infty]$ into $\rho \in [0,\infty]$.
The resulting smoothed mapping for the GZA for
$\alpha=21/13$ and $\gamma=-2$ is shown in Fig.
\ref{scmapping}. 

It turns out that
for the GZA, even after smoothing,
 we can give analytical expressions for the
PDF simply by  using Eq[\ref{nlpdf}] with:
\bea
\de_L &=& \alpha {\rho^{1/\alpha}-1\over{\rho^{\gamma/6+1/\alpha}}}, \\
\left|{d\de_L\over{d\de}}\right| &=&
{\rho^{-1/\alpha-\gamma/6}-
{\gamma\over{6}}~ \de_L\over{\rho}}.
\eea
For example, for the evolved (smoothed) one-point PDF as a function
of $\rho$, we have:
\bea
P(\rho) &=& {1\over{N}}
~~\exp{\left(-
\left[\alpha {\rho^{1/\alpha}-1\over{\rho^{\gamma/6+1/\alpha}}}
\right]^2/(2\sigma^2)
\right)}
~ \times\\ &\times&
\rho^{-1/\alpha-\gamma/6-2}
\left(1-
{\gamma\over{6}}~ \alpha [\rho^{1/\alpha}-1] \right), \nn
\label{GZApdf}
\eea
for Gaussian inital conditions. These sort of compact relations were previously
described
 by Protogeros \& Scherrer (1997). Fig. \ref{fpdf} shows a comparison
of the above PDF to the PDF from simulations and from the SC model.

The velocity smoothing can be obtained from equation (\ref{thetas}) and
equation (\ref{thetagza}):
\beq
\theta = -f_\Omega  \,\, \de_L \, \rho^{1/\alpha+\gamma/6} 
= -f_\Omega \,\, \de_L \, \left| 1 -{\delta_L\over{\alpha}}\right|^{-1}
\label{thetagzas}
\eeq

Note that although the SC and the GZA give a good approximation
to gravitational dynamics for small $\delta$, it is likely and expected 
that these approximations break at some stage for large fluctuations.
As mentioned before, this break is not likely to  be important
for our present purposes, as
the statistics of the absorption features in the 
\lya\ forest  are dominated by the small fluctuations.

\end{document}